\documentclass[fleqn,10pt]{wlscirep}
\usepackage[utf8]{inputenc}
\usepackage[T1]{fontenc}
 \usepackage{relsize}
 \usepackage{txfonts}
 \usepackage{pdfpages}

\title{Impact of individual actions on the collective response of social systems}

\author[1]{Samuel Martin-Gutierrez}
\author[1]{Juan C. Losada}
\author[1,*]{Rosa M. Benito}
\affil[1]{Grupo de Sistemas Complejos, Escuela T\'ecnica Superior de Ingenier\'ia Agron\'omica, Alimentaria y de Biosistemas, Universidad Polit\'ecnica de Madrid, Av. Puerta de Hierro, 2, 28040, Madrid, Spain}

\affil[*]{rosamaria.benito@upm.es}

\begin{abstract}
In a social system individual actions have the potential to trigger spontaneous collective reactions. The way and extent to which the activity (number of actions$-A$) of an individual causes or is connected to the response (number of reactions$-R$) of the system is still an open question. We measure the relationship between activity and response with the distribution of efficiency, a metric defined as $\eta=R/A$. Generalizing previous results, we show that the efficiency distribution presents a universal structure in three systems of different nature: Twitter, Wikipedia and the scientific citations network. To understand this phenomenon, we develop a theoretical framework composed of three minimal statistical models that contemplate different levels of dependence between $A$ and $R$. The models not only are able to reproduce the empirical activity-response data but also can serve as baselines or null models for more elaborated and domain-specific approaches.
\end{abstract}
\begin{document}

\flushbottom

\def\A{\rule{0pt}{1.5ex}A}
\def\R{\rule{0pt}{1.5ex}R}

\def\gf{\mathlarger{\mathlarger{\mathlarger{\gammaup}}}}

\maketitle
%
%
\thispagestyle{empty}


\section*{Introduction}

Due to humans' social nature, the actions of individuals hold the potential to trigger spontaneous collective reactions, leading to complex dynamics. In order to understand human collective behavior, it is necessary to find the laws that relate the individual actions to the collective response of social systems.

This topic has received considerable attention and has been approached from several perspectives \cite{PhysRevE.99.022313, 1903.06588, Muchnik2013, Rybski12640}. From diffusion on networked systems, a field which studies the spread of diseases or information and the emergence of cascading phenomena \cite{ZHANG20161,RevModPhys.87.925,MASUDA20171} to {\it virality}, a property of certain pieces of information that generate a wide response in social systems \cite{doi:10.1287/mnsc.2015.2158, PhysRevE.84.046116, Weng2013}. Other works focus on the Influence Maximization problem, taking advantage of the diffusion mechanisms to find a set of individuals that maximize the response \cite{Kempe:2003:MSI:956750.956769, Zhang2016, 8295265}. Alternatively, the field of control theory aims to steer the collective behavior of a system by controlling the activity of a few individuals \cite{RevModPhys.88.035006, Cremonini2017}.

Our goal in this work is to develop a theoretical framework that relates the number of actions performed by an actor (an agent or individual) embedded in a social system; that is, her activity ($A$), and the number of reactions that these actions trigger in her peers, or response ($R$).
To relate these two magnitudes we generalize the efficiency metric ($\eta = \frac{R}{A}$), introduced by Morales et al. in the context of Twitter \cite{morales2014efficiency}, to other social systems.

We follow a well established modeling approach in social physics: explain the macroscopic properties of the system assuming the simplest microscopic interactions between the actors to extract the most fundamental laws \cite{Iacopini2019,
Williams2019,
Zeng2019,
Candia2019,
Allard2017,
Tamarit8316, 
Newman2010}.
The macroscopic property in which we focus is the distribution of efficiency. We have used this metric to analyze three kinds of social systems of different nature: social networks, collaborative networks and citations networks. 
In particular, we have worked with 14 Twitter conversations around different issues in Spain, Turkey, Palestine, Argentina and Colombia, the editions of the English Wikipedia and the scientific citations data of authors from 14 different countries extracted from the Web of Science. 

In Twitter, the activity is the number of original messages posted by a user and the response of the system is the number of retweets received by that user. Another magnitude used in our analysis is the response to single actions ($r$). In Twitter $r$ would be the number of retweets obtained by a single tweet. In the scientific citations network, $A$ is the number of publications of an author and $R$ the number of citations obtained. The variable $r$ in this case is the number of citations obtained by one paper. In the context of the Wikipedia collaboration network, we consider $A$ as the aggregated number of editions performed by a particular user in any Wikipedia page. The corresponding $R$ is the number of editions made by other users in her personal user page. These editions can be considered as messages directed to that particular user. In this case there is no data for the response to a single edition. Therefore, we have defined $r$ as the number of editions made on the pages of users whose activity is $A = 1$. 

We have found that the efficiency distribution in these three systems has a universal structure with small differences between the datasets, which may indicate the existence of a general mechanism governing the $A-R$ relationship. To reveal that mechanism we have developed three domain-independent minimal statistical models. Taking a parsimonious approach, we start from the most naive model and progressively consider more sophisticated theories with increasingly complex levels of dependence between $R$ and $A$. The models are the Independent Variables model (InV), the Identical Actors model (IdA) and the Distinguishable Actors model (DiA). In the InV model the response of the system is independent with respect to the activity of the individual. In the IdA model, the response of the system depends on the activity of the individual, but the system is agnostic with respect to the individual that stimulates it. Finally, in the DiA model the response is determined not only by the activity of the individual, but also by her features. The models are general because no assumption is made about the particular characteristics of the system or its components.

\section*{Results}

\subsection*{Distribution of efficiency}

The efficiency metric is defined as the quotient between collective response $R$ and individual activity $A$:

\begin{equation}
\eta = \frac{R}{A}
\label{eq:efficiency}
\end{equation}

It can be considered as a proxy for how efficient an individual is at triggering reactions in her peers or as a measure of the system's  inertia to react to the stimuli of the individual. The higher the individual's efficiency, the lower the system's inertia.

\begin{figure}[htbp]
\centering
\includegraphics[width=0.5\linewidth]{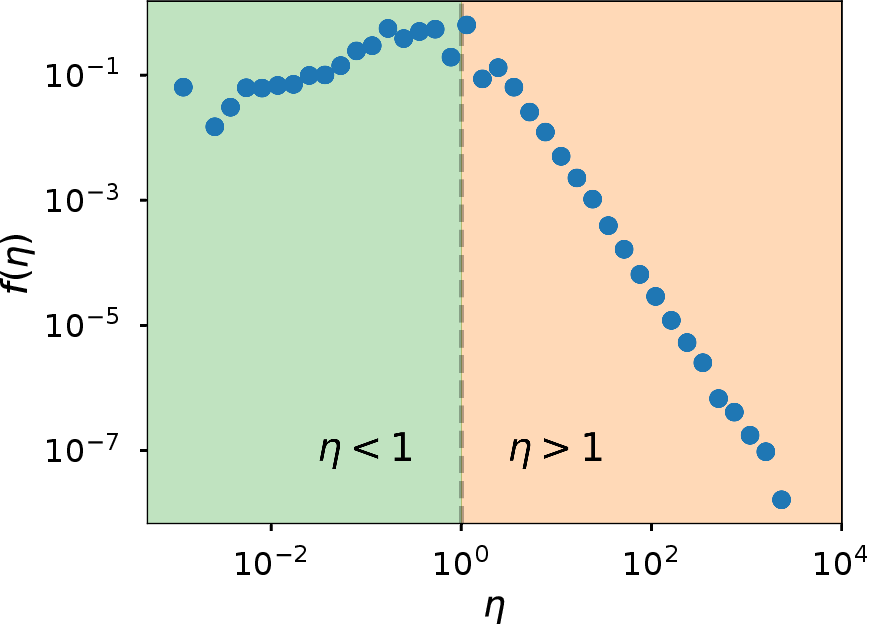}
\caption{
Example of efficiency distribution where the two distinct behaviors that are manifested to each side of the point $\eta=1$ can be appreciated. The data corresponds to the Twitter conversation around the 2015 Spanish General Elections.}
\label{fig:eff_distr_exmpl}
\end{figure}

Our work is focused on the efficiency distribution, an example of which is presented in figure \ref{fig:eff_distr_exmpl}. It is characterized by a concave shape with two distinct tendencies for $\eta<1$ (an individual gets less than one reaction per action) and $\eta>1$ (an individual triggers more than one reaction per action). In the work by Morales et al. \cite{morales2014efficiency}, they used the Independent Cascade (IC) model on the Twitter follower network to reproduce the empirical distribution of user efficiency and showed that the shape of the distribution was universal for Twitter conversations. However, several questions were left open and some of the empirical results lacked a comprehensive explanation. In particular, they reported evidence for the independence of the efficiency distribution with respect to the functional form of the activity distribution and, from that, conjectured that communication patterns are not dependent on the way users post original messages; that is, that collective response is independent of individual activity.

In this work we go one step further and present evidence for the universality of the structure of the efficiency distribution in two other social systems. We also present the three aforementioned statistical models to provide a comprehensive description of the nature of the efficiency distribution and show the extent to which the activity of the individuals and their particular features influence the response of the system.

\subsection*{Description of the models} 

We have calculated the theoretical distributions of efficiency with three different methodologies: 
Monte-Carlo (MC) simulation, direct computation with discrete probability distributions and derivation of an analytical expression.

Once the basic mechanism of the model is laid out, MC simulation allows a direct implementation of the model's assumptions. Thus, we use it to compare model and empirical data as well as to verify the results of the other methodologies.

To directly compute the efficiency distribution with the discrete joint probability distribution $p(R,A)$ we follow the method described in the Methods section (equations \eqref{eq:jp_2_eff_discr_CDF} and \eqref{eq:jp_2_eff_discr_PDF}). The resulting efficiency distribution is asymptotically exact in the sense that, since the support for the distributions of $A$ and $R$ is $\mathbb{N}$, an infinite number of terms would be required to actually obtain exact results, but larger values of $A$ and $R$ have increasingly smaller probabilities, carrying progressively lower weight on the computation and enabling the results to converge for a finite number of terms.

The analytical calculation of the efficiency distribution has been carried out for the InV and IdA models by considering $A$ and $R$ as continuous random variables. Taking into account the definition of efficiency given by \eqref{eq:efficiency} we derive an expression for the probability density function (PDF) of efficiency using the joint probability distribution $\varphi(R,A) = \varphi(\eta A,A)$ (see 
section 2 of
the Supplementary Information):

\begin{equation}
  f(\eta)=\begin{cases}
                \int_{R_m/\eta}^{\infty} \varphi(\eta A,A) A dA \quad \text{if} \quad \eta \leq \frac{R_m}{A_m}\\
                \int_{A_m}^{\infty} \varphi(\eta A, A) A dA \quad \text{if} \quad \eta > \frac{R_m}{A_m}\\
            \end{cases}
\label{eq:eff_anal_general}
\end{equation}

Where $A_m, R_m>0$ are the minimum values of $A$ and $R$. In our case, $A_m = R_m = 1$ for every dataset. 
It is worth noting that the two branches of $f(\eta)$ in equation \eqref{eq:eff_anal_general} correspond to the two characteristic tails of the efficiency distribution.

\subsubsection*{Independent Variables model}

In the InV model $A$ and $R$ are considered independent variables with probability distributions $p(A)$ and $p(R)$. 

A Monte-Carlo simulation can be computed as follows: In a system with $N$ individuals indexed by $i=1,2,\dots, N$, store the empirical data of activity and response in two vectors $\vec{A}$ and $\vec{R}$ such that component $i$ of vector $\vec{A}$ corresponds to the same individual as component $i$ of vector $\vec{R}$. Next, shuffle each of them independently, such that the correlations that may have been present when each couple $(A_i,R_i)$ corresponded to the same individual vanish. The randomized versions of the vectors, $\vec{A}_{rnd}$ and $\vec{R}_{rnd}$, hold the same values as the originals but with the order of the elements randomly altered. Finally,  the efficiency vector $\vec{\eta}_{rnd} = \vec{R}_{rnd} / \vec{A}_{rnd}$ is used to compute the efficiency distribution according to the InV model.

Since $A$ and $R$ are considered independent, their discrete joint probability distribution is $p(R,A)=p(R)p(A)$. The PDF of efficiency can be obtained by plugging this expression  in \eqref{eq:jp_2_eff_discr_CDF} and \eqref{eq:jp_2_eff_discr_PDF} of Methods. However, for this model we have left out the results of the discrete methodology because we have derived an exact analytical expression.

For the analytical computation of the InV model we consider $A$ and $R$ as continuous variables with PDFs $f_A(A)$ and $f_R(R)$. Their joint probability distribution can be written as $\varphi(R,A)=f_A(A)f_R(R)$. Plugging this in \eqref{eq:eff_anal_general} we obtain:

\begin{equation}
f^{InV}(\eta)=\begin{cases}
                \int_{R_m/\eta}^{\infty} f_R(\eta A)f_A(A) A dA \quad \text{if} \quad \eta \leq \frac{R_m}{A_m}\\
                \int_{A_m}^{\infty} f_R(\eta A)f_A(A) A dA \quad \text{if} \quad \eta > \frac{R_m}{A_m}\\
            \end{cases}
\label{eq:eff_distrib_InV_res}
\end{equation}

This expression provides an explanation for a key result presented in \cite{morales2014efficiency}, where Morales et al. show that the right tail of the efficiency distribution remains unaltered when the activity distribution is modified. To reach that result, let us assume that $f_R(R)\propto R^{-\gamma^{}_R}$. This power law distribution was used in \cite{morales2014efficiency} as well as in other works to model the distribution of retweets \cite{borondo2012characterizing}, scientific citations \cite{doi:10.1137/070710111} and incoming editions in Wikipedia \cite{Muchnik2013}. Then, the right tail ($\eta > \frac{R_m}{A_m}$) of the PDF shown in \eqref{eq:eff_distrib_InV_res} can be written as:

 \begin{equation}
 f^{InV}(\eta) \propto \eta^{-\gamma^{}_R} \int_{A_m}^{\infty} A^{1-\gamma^{}_R}f_A(A) dA = E_A[A^{1-\gamma^{}_R}] \eta^{-\gamma^{}_R} 
 \Rightarrow
 f^{InV}(\eta) \propto f_R(\eta)
 \label{eq:eff_distr_tail_res}
 \end{equation}

Where $E_A[\cdot]$ is the expected value with respect to the activity distribution. Therefore, when $f_R(R)\propto R^{-\gamma^{}_R}$, the right tail of the efficiency distribution is proportional to $\eta^{-\gamma^{}_R}$. That is, in addition to being independent of the activity distribution, its shape is completely determined by the exponent of the response distribution.

To apply the analytical computation of the efficiency distribution for the InV model to empirical data we have fit the empirical distributions of $A$ and $R$ to a power law with exponential cutoff (or truncated power law) using the powerlaw python module \cite{10.1371/journal.pone.0085777}. The functional form of this distribution is the following:

\begin{equation}
f(x) =  \frac{\lambda^{1-\alpha}}{\Gamma(1-\alpha,\lambda x_{min})} x^{-\alpha}e^{-\lambda x}
\end{equation}

Where $\Gamma(s, x)$ is the upper incomplete gamma function. 
The resulting fits for $f_A(A)$ and $f_R(R)$ for every dataset are presented in the Supplementary Information (SI). When the PDFs of activity and response are power laws with exponential cutoff, the PDF of efficiency adopts the following form:

\begin{equation}
f^{InV}(\eta)=\begin{cases}
                g(\eta) \Gamma(2-\alpha_{\R}-\alpha_{\A},(\lambda_{\R} \eta+\lambda_{\A})\frac{R_m}{\eta}) 
                \quad \text{if} \quad \eta \leq \frac{R_m}{A_m}\\
                g(\eta) \Gamma(2-\alpha_{\R}-\alpha_{\A},(\lambda_{\R} \eta+\lambda_{\A})A_m)
                \quad \text{if} \quad \eta > \frac{R_m}{A_m}\\
            \end{cases}\\
\label{eq:inp_eff_distr_res}
\end{equation}

With

\begin{equation}
 g(\eta) = C (\lambda_{\R}\eta+\lambda_{\A})^{(\alpha^{}_R+\alpha^{}_A-2)}\eta^{-\alpha^{}_R}
\end{equation}

and

\begin{equation}
C = 
\frac{\lambda_R^{1-\alpha^{}_R}}{\Gamma(1-\alpha^{}_R,\lambda^{}_RR_m)}
\frac{\lambda_A^{1-\alpha^{}_A}}{\Gamma(1-\alpha^{}_A,\lambda^{}_AA_m)}
\end{equation}

\subsubsection*{Identical Actors model}

A natural extension to the InV model is to consider that the response of the system depends on the activity of the individual. To carry out this extension in a parsimonious way, we realize that the stimuli to which the system reacts occur in a discrete fashion, so we can assume that it reacts to each action (a tweet, a scientific publication, an edition on Wikipedia, etc.) individually, as if they were isolated events. Then, while in the InV model the {\it aggregate} response of the system was independent of the aggregate activity of the actor, in the IdA model the {\it partial} response of the system to each single action is independent of the actor. But, as the aggregate response of the system to the activity of an individual is the sum of the partial responses to each of her $A$ actions, a dependence between $R$ and $A$ is induced.

To formalize this idea we introduce the new variable $r$ as the response of the system to a single action by any individual. This random variable follows the same distribution $p(r)$ for all the actors. The aggregate response $R$ associated to an actor that performed $A$ actions and triggered partial responses $\{r_1,r_2,\dots,r_A\}$ is $R=\sum_{j=1}^A r_j$. The dependence of $R$ on $A$ resides on the number of terms of this sum.


To perform a Monte-Carlo simulation of the IdA model, we first fit the $p(r)$ with the hybrid methodology detailed in the SI and $p(A)$ to a discrete truncated power law (see the SI for the results). Then, we generate a set of individuals whose activity is assigned according to $p(A)$. The responses for each of the $A$ actions of an individual is randomly generated with $p(r)$ and then aggregated to obtain her $R$. The efficiency according to this model is directly computed from the $(R,A)$ tuple associated to each actor.

To get the efficiency distribution of the IdA model from the discrete $p(R,A)$ distribution, we start with the conditional discrete probability distribution of $R$ given an activity $A$, which is computed as the $A-fold$ discrete convolution of $p(r)$ with itself:

\begin{equation}
p(R|A) = p(r_1)*p(r_2)*\cdots*p(r_A) = p(r)*p(r)*\cdots*p(r) = p^{*A}(r)
\end{equation}

Then, the joint probability distribution can be obtained as: 

\begin{equation}
p(R,A) = p(R|A)p(A) = p^{*A}(r) p(A)
\label{eq:jp_act_conv_distr_res}
\end{equation}

The efficiency PDF is obtained by plugging \eqref{eq:jp_act_conv_distr_res} in \eqref{eq:jp_2_eff_discr_CDF} and \eqref{eq:jp_2_eff_discr_PDF}. The $p(r)$ and $p(A)$ distributions used in this methodology are the same as those used in the Monte-Carlo simulations.

To carry out the previous computations with infinite precision we would need an infinite number of values for the $p(r)$, $p(A)$ and $p(R,A)$ distributions. To be able to perform the numerical computations, we have used distributions that are bounded at a certain value and we have verified that further increasing the number of values employed do not affect the results. The cut-off values used for the three systems considered are shown in table \ref{tab:cut_off}.


\begin{table}[htbp]
 \begin{center}
\begin{tabular}{|l|c|c|c|}
\hline
{\bf System} & \boldmath$r_{max}$ & \boldmath$A_{max}$ & \boldmath$R_{max}$  \\ \hline
Twitter & $10^6$ & $3\times10^4$ & $5\times10^4$ \\ \hline
Wikipedia & $3\times10^5$ & $3\times10^4$ & $5\times10^4$ \\ \hline
Citations & $3\times10^5$ & $2\times10^4$ & $3\times10^4$ \\ \hline
\end{tabular}
\end{center}
\caption{
Cut-off values used to perform the numerical computations for the IdA model.}
\label{tab:cut_off}
\end{table}

An analytical expression for the efficiency distribution of the IdA model can be derived when $p(r)$ is modeled as a power law ($p(r)\propto r^{-\gamma^{}_r}$). For this approximation, the activity distribution $p(A)$ has been modeled as a power law with exponent $\gamma_{\A}$, a usual approach in the literature \cite{Muchnik2013, borondo2012characterizing}. The corresponding fits are shown in the SI and the resulting expression for the PDF of efficiency is:

\begin{align}
f^{IdA}(\eta) = \begin{cases}
\begin{cases}
                C \frac{1}{2-\gamma^{}_r} \eta^{\gamma^{}_A - 3} E_{\frac{\gamma^{}_A-1}{2-\gamma^{}_r}}(\eta^{-1}) 
                & \quad \text{if} \quad \eta \leq 1\\[5 pt]
                C \frac{1}{2-\gamma^{}_r} \eta^{-\gamma^{}_r} E_{\frac{\gamma^{}_A-1}{2-\gamma^{}_r}}(\eta^{1-\gamma^{}_r}) 
                & \quad \text{if} \quad \eta > 1\\
            \end{cases} & \quad \text{when} \quad \gamma^{}_r < 2\\[30 pt]
\begin{cases}
                C \eta^{\left(-1-\frac{(-1-\gamma^{}_r)(1-\gamma^{}_A)}{2-\gamma^{}_r}\right)}
                \frac{1}{\gamma^{}_r - 2} 
                \gf \left( \frac{3-\gamma^{}_r-\gamma^{}_A}{2-\gamma^{}_r} , \eta^{-1} \right)
                & \quad \text{if} \quad \eta \leq 1\\[5 pt]
                C \eta^{\left(-1-\frac{(-1-\gamma^{}_r)(1-\gamma^{}_A)}{2-\gamma^{}_r}\right)}
                \frac{1}{\gamma^{}_r - 2} 
               \gf \left( \frac{3-\gamma^{}_r-\gamma^{}_A}{2-\gamma^{}_r} , \eta^{1-\gamma^{}_r} \right)
                & \quad \text{if} \quad \eta > 1\\
            \end{cases} & \quad \text{when} \quad \gamma^{}_r > 2 
 \end{cases}
\label{eq:idp_eff_pdf_anal_res}
\end{align}

Where $E_n(\cdot)$ is the generalized exponential integral, $\gf(s,x)$ the lower incomplete gamma function and $C$ the following normalization constant:

\begin{equation}
C = \frac{(\gamma^{}_r-1)(\gamma^{}_A-1)}{1+(1-\gamma^{}_A)\Gamma(1-\gamma^{}_A,1)}
\end{equation}

\subsubsection*{Distinguishable Actors model}

In the DiA model the actors are distinguishable, meaning that the system is sensitive to the individual who makes the action and reacts in a different manner depending on her particular features. 

This idea can be formalized by considering that the probability distribution of response to single actions depends on the features of the individual that performs the action, summarized in a vector $\vec{s}$. The distribution of aggregate response $R$ of the system is computed as the $A-fold$ convolution of the $p(r|\vec{s})$ distribution with itself:
 
 \begin{equation}
 p(R|A,\vec{s}) = p^{*A}(r|\vec{s})
 \end{equation}
 
If $\{s_1,s_2,\dots,s_N\}$ are the components of the feature vector (assume the features are independent discrete variables), the discrete joint probability distribution $p(R,A)$ is obtained as follows:

\begin{equation}
p(R,A) = \sum_{s_1} \cdots \sum_{s_N} p^{*A}(r|\vec{s}) p(A) p(\vec{s})
\end{equation}

Finally, $p(R,A)$ can be used to compute the efficiency distribution with \eqref{eq:jp_2_eff_discr_CDF} and \eqref{eq:jp_2_eff_discr_PDF}.

A key point is to find the conditional probability distribution $p(r|\vec{s})$ that characterizes the relationship between the features $\vec{s}$ of the individual and the response $r$ of the system to her actions. Unfortunately, this task is not trivial in most cases. In the case of the citation network the literature shows that there are many and varied factors that determine the citation counts of publications \cite{BORNMANN201211}, from the quality of the manuscript, to the field of research, the cited references or the reputation of the authors and their institutions. With respect to Wikipedia, some factors that could determine the response to a user could be the topics she is more active on, the age of her user account or her main role (some users may be focused on editing articles, others on moderating discussion pages, etc.). 

Among the systems under study we have focused on Twitter, where we have chosen the number of followers $F$ of a user as a proxy of her ability to trigger a response, since the follower layer is the substrate through which the retweets are spread \cite{bakshy2011everyone, BORONDO201590}.

In order to establish the relationship between an individual's features and the response of the system, we have relied on the Independent Cascade (IC) diffusion model. We have formalized the IC model by means of the binomial distribution and a set of assumptions based on empirical evidence (see Methods), obtaining the following expression for the response distribution to single actions conditioned on the number of followers ($F$) of the individual:

\begin{equation}
p(r|\vec{s}) = p(r|F) = B(r;F,p_{inf})
\label{eq:dip_cond_r_F_res}
\end{equation}

Where $B(x;n,p)$ is a binomial distribution. The discrete joint probability distribution for $A$ and $R$ is given by:

 \begin{equation}
 p(R,A) = p(A) \sum_{F=0}^\infty B(R;AF,p_{inf}) p(F)
 \label{eq:dip_jp_ra_res}
 \end{equation}
 
The PDF of efficiency is obtained by plugging  \eqref{eq:dip_jp_ra_res} in \eqref{eq:jp_2_eff_discr_CDF} and \eqref{eq:jp_2_eff_discr_PDF}.
 
Notice that $F$ is the only component of the feature vector $\vec{s}$ of the individual. The infection probability parameter $p_{inf}$ has been considered constant and equal for every individual and has been determined by Maximum Likelihood Estimation (MLE) of the $p(r)$ distribution. The discrete computation of the DiA model also requires a fit for the $p(F)$ distribution, which was performed with the hybrid methodology detailed in the SI. The $p(A)$ was fit to a discrete truncated power law.

A Monte-Carlo simulation of the DiA model can be performed as follows: Generate a set of individuals with a random number of followers $F \sim p(F)$ and a random activity $A \sim p(A)$. Then, for each action $j$ ($j=1,2,\dots,A$) performed by an individual, the partial response of the system $r_j$ is computed with \eqref{eq:dip_cond_r_F_res} and the aggregate response with $R=\sum_{j=1}^A r_j$. 

For this model, we have found that an analytical derivation of the PDF of efficiency is too cumbersome to be tackled.

\begin{figure}[htbp]
\centering
\includegraphics[width=0.8\linewidth]{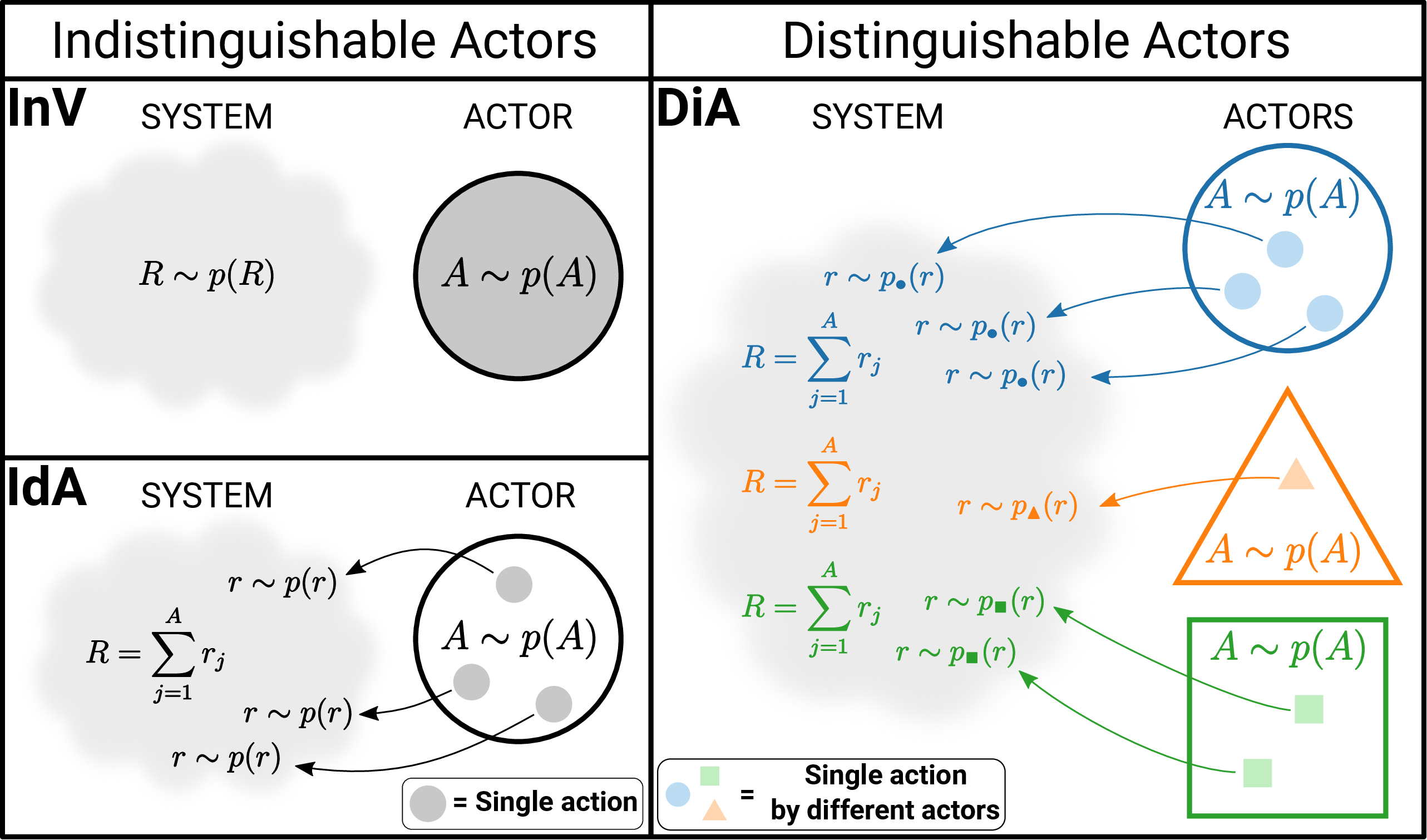}
\caption{
Diagram that summarizes the main characteristics of the three models: Independent Variables (InV), Identical Actors (IdA) and Distinguishable Actors (DiA).}
\label{fig:diagram_models}
\end{figure}

To conclude this section, we summarize the main features of the three developed models in figure \ref{fig:diagram_models}. The models can be classified taking into account two properties: the distinguishability of the actors and the dependence of $R$ with respect to $A$.  Concerning the distinguishability of the actors, we have on one side the InV and IdA models, where the actors are considered identical, and on the other side the DiA model, where the particular features of the actors are taken into account. Regarding the $A-R$ dependence, we have on one side the InV model, in which $R$ and $A$ are independent variables, and on the other side, the IdA and DiA models, where $R$ depends on $A$ because the aggregate response $R$ is the sum of the partial responses $r$ to each individual action.


\subsection*{Application of the models to empirical data}

The models presented in the previous section have been tested in three different systems: the scientific citations network, Twitter and Wikipedia. 
See the Supplementary Information (SI) for a detailed description of the datasets.
In this section we analyze the models' performance in each of them.

\subsubsection*{Independent Variables model}

\begin{figure}[htbp]
\centering
\includegraphics[width=\linewidth]{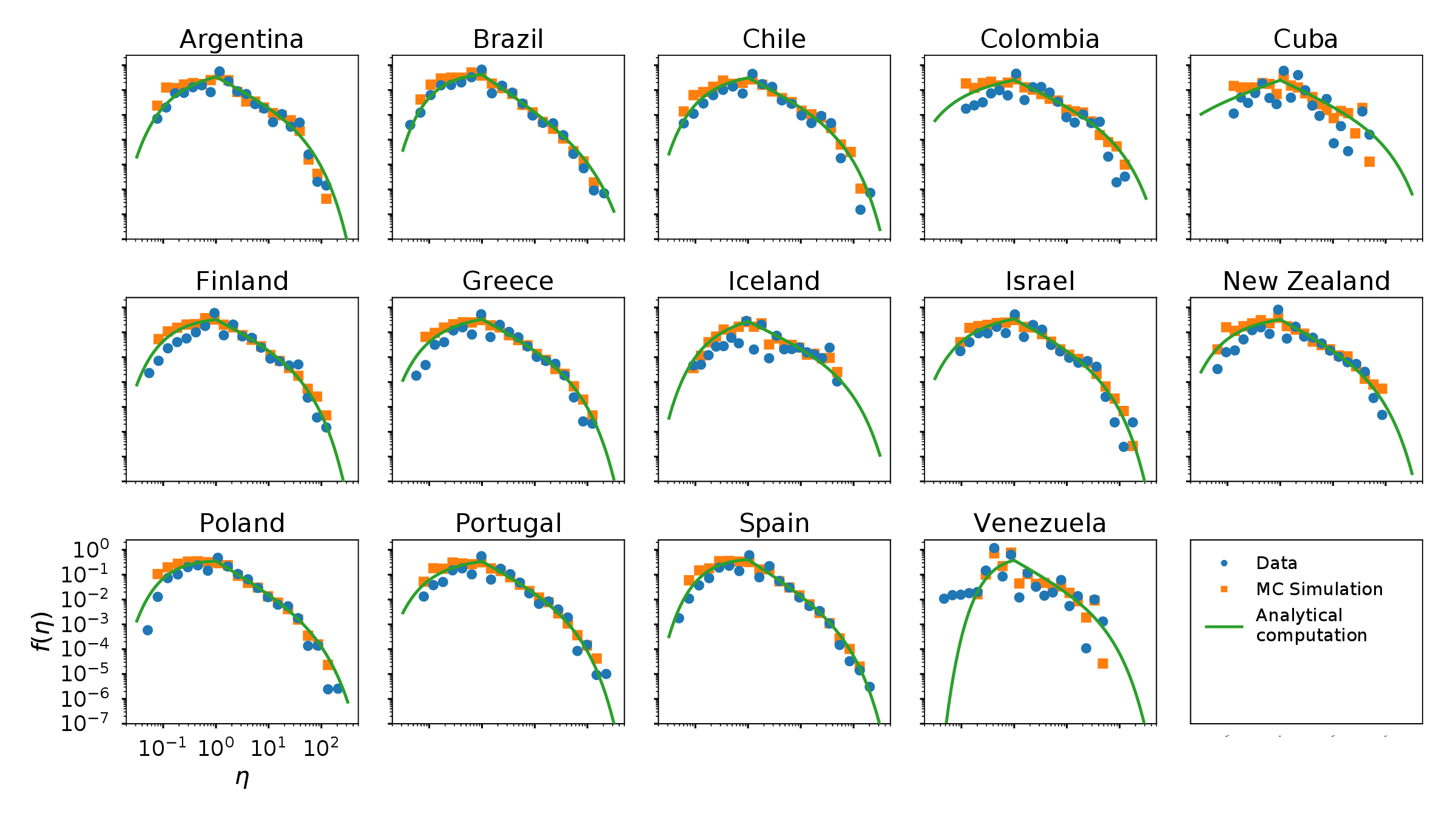}
\caption{
Efficiency distributions corresponding to the InV model applied to the scientific citations dataset. The plots show the empirical efficiency distribution (blue dots), the Monte-Carlo simulation (orange squares) and the analytical expression of \eqref{eq:inp_eff_distr_res} (green line).}
\label{fig:sci_inp_eff}
\end{figure}

\begin{figure}[htbp]
\centering
\includegraphics[width=\linewidth]{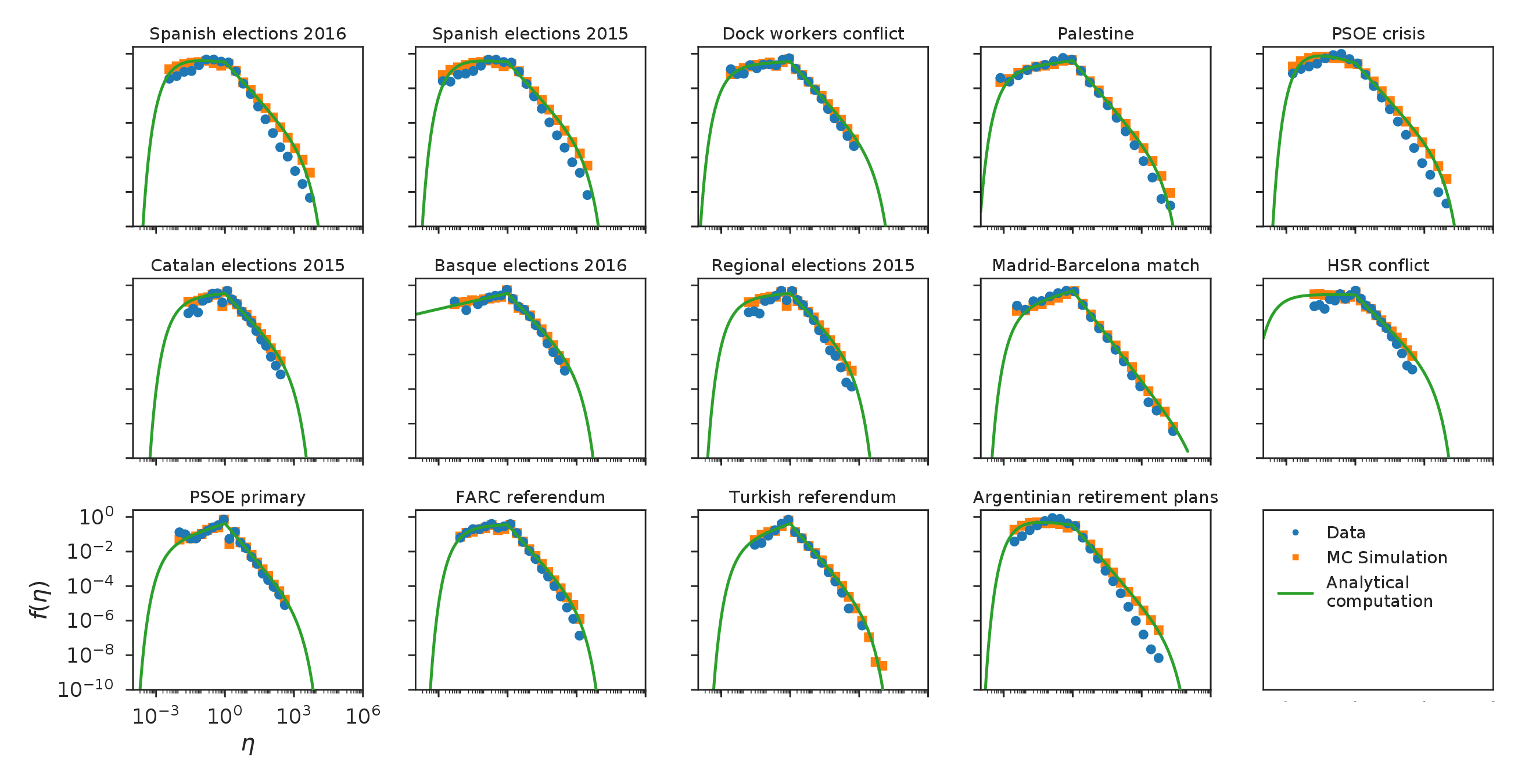}
\caption{
Same as figure \ref{fig:sci_inp_eff} for the InV model applied to the Twitter datasets.}
\label{fig:twt_inp_eff}
\end{figure}

\begin{figure}[htbp]
\centering
\includegraphics[width=.5\linewidth]{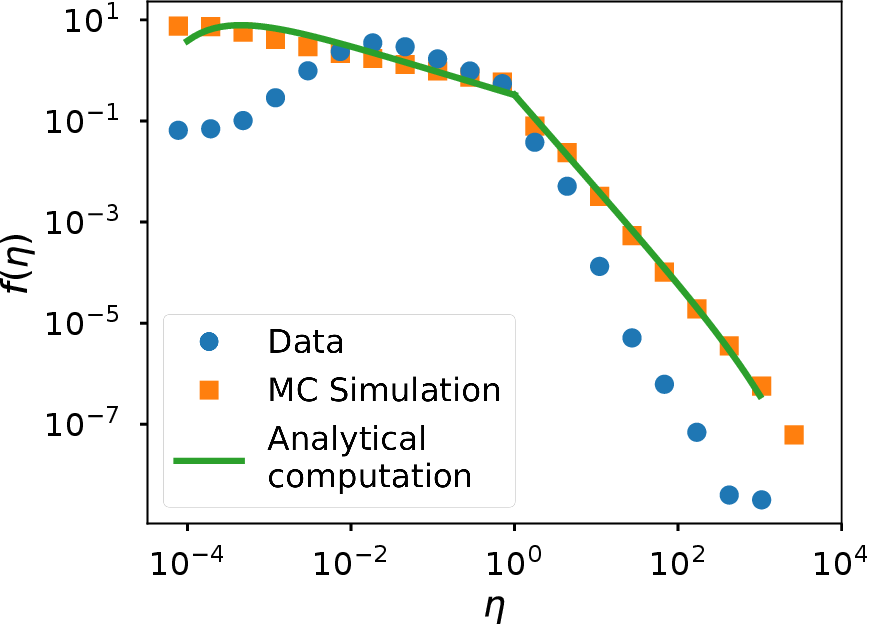}
\caption{
Same as figure \ref{fig:sci_inp_eff} for the InV model applied to the Wikipedia dataset.}
\label{fig:wiki_inp_eff}
\end{figure}

In figures \ref{fig:sci_inp_eff},  \ref{fig:twt_inp_eff} and \ref{fig:wiki_inp_eff} we present the empirical and theoretical (Monte-Carlo simulation and analytical expression) efficiency distributions according to the InV model for scientific citations, Twitter and Wikipedia respectively. In the case of the scientific citations datasets the model adequately reproduces the efficiency distribution in most cases; in particular, we obtain very good fits for the datasets of Brazil and Spain, which are the largest (see SI). 

As can be seen in figure \ref{fig:twt_inp_eff}, the InV model captures the general shape of the Twitter efficiency distributions. Nevertheless, on closer examination we see that for several datasets the model does not fully agree with the empirical data, especially in the right tails (see Spanish elections 2015/2016, PSOE crisis or Argentinian retirement plans for example), although for a few others the concordance is quite good (see Madrid-Barcelona match or PSOE primary). Besides, the analytical expression of \eqref{eq:inp_eff_distr_res} shows a very good agreement with the Monte-Carlo computation in every case, validating the approximations adopted for the analytical derivation.

The efficiency distribution of the InV model presents a partial agreement with the Wikipedia data in the left tail (see figure \ref{fig:wiki_inp_eff}),  but a deviation can be appreciated in the lowest values of efficiency and in the right tail. On the other hand, the analytical expression of \eqref{eq:inp_eff_distr_res} shows a good concordance with the Monte-Carlo simulation. 

 Summing up, the InV model captures the universal structure of the efficiency distribution. Furthermore, it reproduces with good accuracy the empirical data of the scientific citations network. However, we achieve slightly worse agreements for several of the Twitter datasets and obtain a higher discrepancy between model and data for Wikipedia. This difference in performance may arise from the long reaction times of citations and prolonged lifespan of scientific publications (months or years) \cite{Cano1991}, which contrast with the short characteristic times of the interactions in social media  (in the order of minutes or hours) \cite{Zhao:2015:SSP:2783258.2783401, kobayashi2016tideh, 10.1371/journal.pone.0038869}. Another important factor to take into account is that while the actions (and the corresponding reactions) in the case of Twitter and Wikipedia are associated to a specific individual because a tweet or an edition have one author (who is the recipient of the retweets and incoming messages), a scientific paper usually has several authors and its associated citations are assigned to all of them. Moreover, as mentioned above, there are many and varied circumstances that determine the citation counts of publications. Therefore, the independence assumption may work for the scientific citations datasets because the different overlapping factors discussed above mask the dependence between $R$ and $A$.

Conversely, the model-data discrepancies observed in the other systems may emerge because in that case the independence assumption is not fully adequate. To verify this hypothesis, we check if the $A-R$ correlations neglected by the independence assumption affect the quality of the fit given by the InV model: First, we compute the empirical Spearman's rank correlation $\rho_e$ \cite{corder2014nonparametric} between $R$ and $A$. Then, we measure the discrepancy $\Delta^{}_{InV}$ between the InV model and the data. Finally, we test if $\rho_e$ and $\Delta^{}_{InV}$ are positively correlated.

The disagreement between the InV model and the data mostly lie on the extreme values of the right tail of the efficiency distribution, which have very low probabilities. Since the right tail can be approximated by a power law, we define $\Delta^{}_{InV}$ as the difference between the theoretical exponent ($\gamma^{}_{InV}=\gamma^{}_R$ from equation \eqref{eq:eff_distr_tail_res}) and empirical exponent ($\gamma^{}_{emp}$) of that power law:

\begin{equation}
\Delta^{}_{InV} = \gamma^{}_{emp} - \gamma^{}_{InV} = \gamma^{}_{emp} - \gamma^{}_{R}
\label{eq:delta_inv}
\end{equation}

The power law exponent reflects the global trend including the effect of the tail, so the contributions to the error of low probability values are not underestimated. See the SI for more details in this respect.

The relationship between $\Delta_{InV}$ and the empirical correlation $\rho_e$ has been tested by means of a linear regression carried out with the Twitter datasets. This regression (shown in figure S2 of the SI) yields a positive correlation of $r=0.80$, indicating that $\Delta^{}_{InV}$ increases monotonously with $\rho_e$ and corroborating the hypothesis presented above.

\subsubsection*{Identical Actors model}

\begin{figure}[htbp]
\centering
\includegraphics[width=\linewidth]{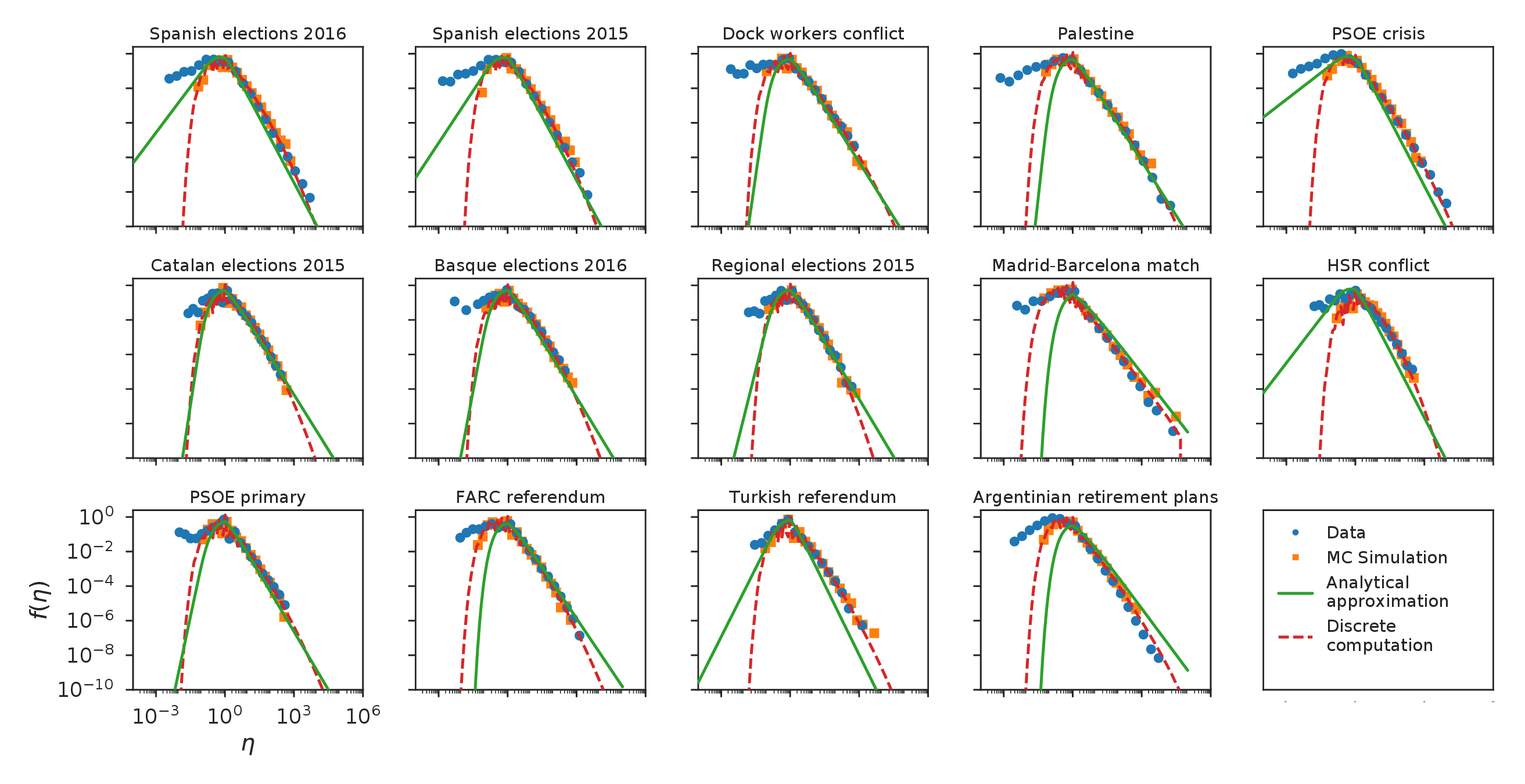}
\caption{
Efficiency distributions corresponding to the IdA model applied to the Twitter datasets. The plots show the empirical efficiency distribution (blue dots), the Monte-Carlo simulation (orange squares), the discrete computation (dashed red line) and the analytical expression of \eqref{eq:idp_eff_pdf_anal_res} (green line).}
\label{fig:twt_idp_eff}
\end{figure}

\begin{figure}[htbp]
\centering
\includegraphics[width=.5\linewidth]{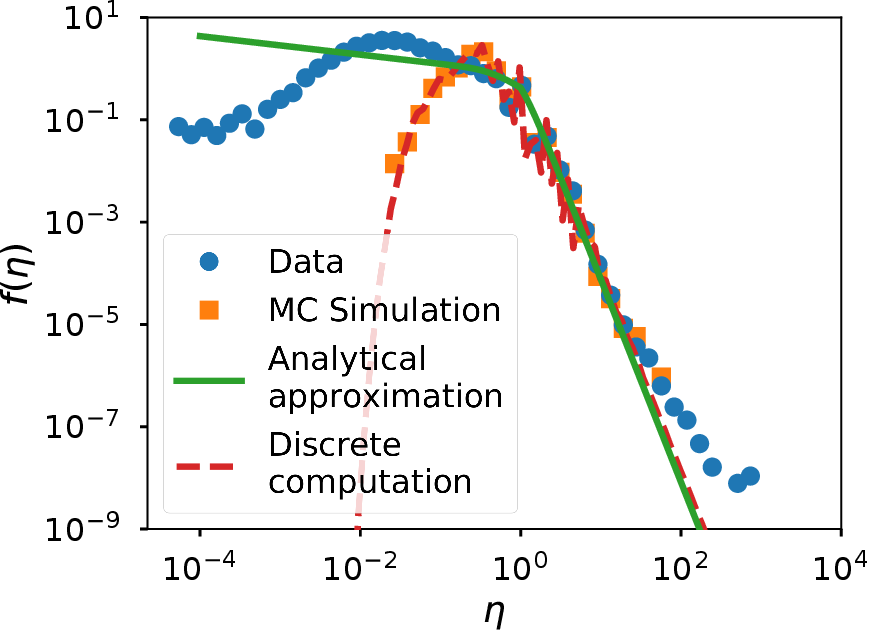}
\caption{
Same as figure \ref{fig:twt_idp_eff} for the IdA model applied to the Wikipedia dataset.}
\label{fig:wiki_idp_eff}
\end{figure}

\begin{figure}[htbp]
\centering
\includegraphics[width=\linewidth]{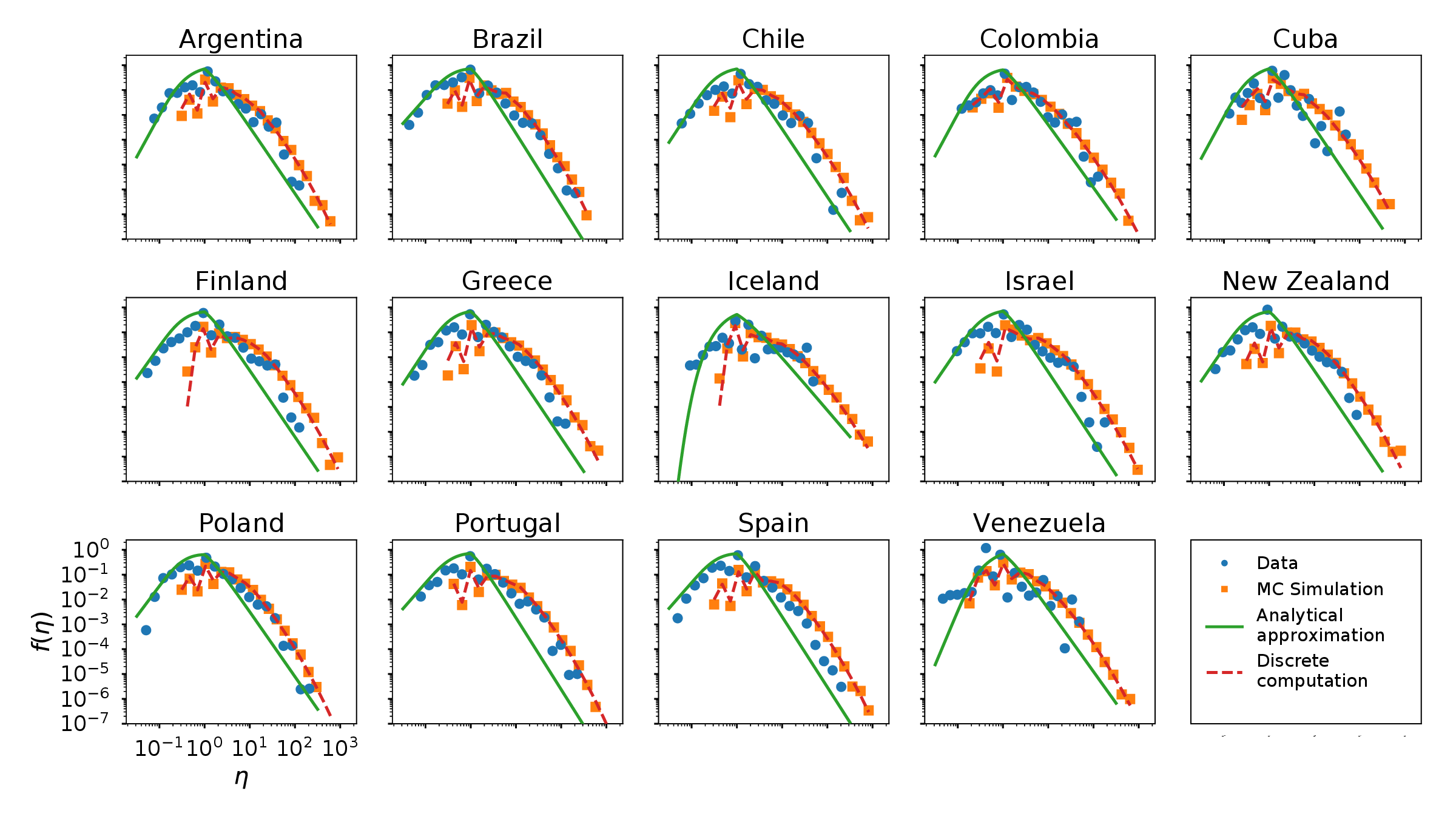}
\caption{
Same as figure \ref{fig:twt_idp_eff} for the IdA model applied to the scientific citations datasets.}
\label{fig:sci_idp_eff}
\end{figure}

Since the $A-R$ correlations seem to be the cause of the discrepancy between the InV model and the data, we expect that the dependence between $R$ and $A$ introduced by the IdA model improves the previous results for Twitter and Wikipedia. In figures \ref{fig:twt_idp_eff}, \ref{fig:wiki_idp_eff} and \ref{fig:sci_idp_eff} we present the efficiency distributions for Twitter, Wikipedia and scientific citations according to the IdA model. 
A clear improvement in the agreement between theory and data can be appreciated on the right tails of all the Twitter datasets shown in figure \ref{fig:twt_idp_eff}. However, there is a higher discrepancy in the left tail of the distribution, which will be discussed below. 

If we compare the results of the model got by the analytical expression given by \eqref{eq:idp_eff_pdf_anal_res} and the discrete computation we find a good correspondence in general. However, there are small deviations that can be explained by the approximations concerning the power-law fit of $p(r)$ (see SI for details).

In figure \ref{fig:wiki_idp_eff} we can see that for Wikipedia the IdA model is also capable of reproducing the right tail of the distribution adequately. The left tail however falls too fast in this model.

Concerning the scientific citations, the IdA model (in figure \ref{fig:sci_idp_eff}) presents a slightly worse agreement in comparison with the InV model. Therefore, the scientific citations network is better characterized by a model in which activity and response are considered independent. The reasons were discussed in the previous section.

In the IdA model, contrasting with the InV model, $R$ and $A$ are correlated; but there is no guarantee that the theoretical correlations match the empirical ones. To verify this, we have carried out a linear regression (see figure S3 of the SI) between the theoretical Spearman's correlation $\rho_t$ (averaged from 300 Monte-Carlo realizations of the IdA model) and the empirical one $\rho_e$ for the Twitter datasets, obtaining an $r^2=0.88$ and the following equation for the line: $\rho_t = (1.30\pm0.14) \rho_e - (0.13\pm0.06)$. As it can be appreciated, there is a significant correspondence between empirical and theoretical correlations: the slope is close to $1$ and the value of the intercept is close to $0$, corroborating the ability of the IdA model to reproduce the correlations of the real data to a reasonable extent.

Summarizing, there is an excellent agreement between the IdA model and the data in the right tail of the efficiency distribution for Twitter and Wikipedia. Besides, the correlations induced by the model present a high correspondence with the empirical correlations between $A$ and $R$. However, the left tail of the efficiency distribution falls faster in the model than in the data. A low efficiency implies that an individual performs many actions obtaining a very low aggregate response  (high $A$ and low $R$), but under the IdA model, which considers that {\it every individual and every action have identical capability} to trigger a response on the system, low efficiencies are very unlikely, as performing many individual actions guarantee at least a moderate aggregate response.

Rather, the empirical evidence suggests that individuals from the left tail of the efficiency distribution ($\eta < 1$) have lower capabilities to trigger a response than those of the right tail ($\eta>1$).  This is backed by the fact that the InV model, which considers response independent from activity, reproduces the left tail of the efficiency distribution with better accuracy than the IdA model. Therefore, we hypothesize that the response associated to actors with lower efficiencies do not depend on their activity (so they should show low $A-R$ correlation), contrasting with the users with higher efficiencies, whose behavior can be characterized with the IdA model and should present high $A-R$ correlation.

\begin{figure}[htbp]
\centering
\includegraphics[width=.5\linewidth]{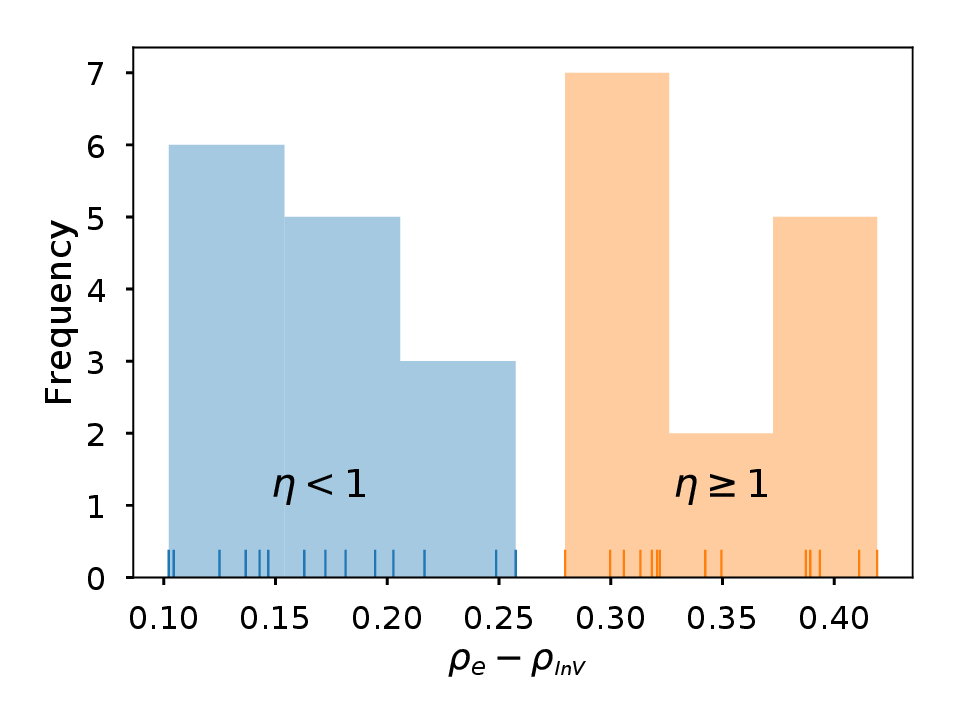}
\caption{
Histogram of the Spearman's rank correlation between $A$ and $R$ for the Twitter datasets computed for individuals with low efficiency ($\eta<1$) and high efficiency ($\eta \geq 1$). The small vertical lines in the bottom of the x axis are a rug plot and represent the value of $\rho_e-\rho_{InV}$ for each individual dataset.}
\label{fig:LH_eff_corr_hists}
\end{figure}

The previous hypothesis can be verified by comparing the Spearman's rank correlation between $A$ and $R$ for individuals with $\eta<1$ and $\eta>1$. However, we should be careful and take into account the artificial correlations induced by performing this filtering. This is achieved by subtracting the correlation associated to the InV model (averaged from 300 realizations). We have performed this computation for the Twitter datasets and, as can be appreciated in figure \ref{fig:LH_eff_corr_hists}, there is a clear difference between both sets of individuals.

\subsubsection*{Distinguishable Actors model}

\begin{figure}[htbp]
\centering
\includegraphics[width=\linewidth]{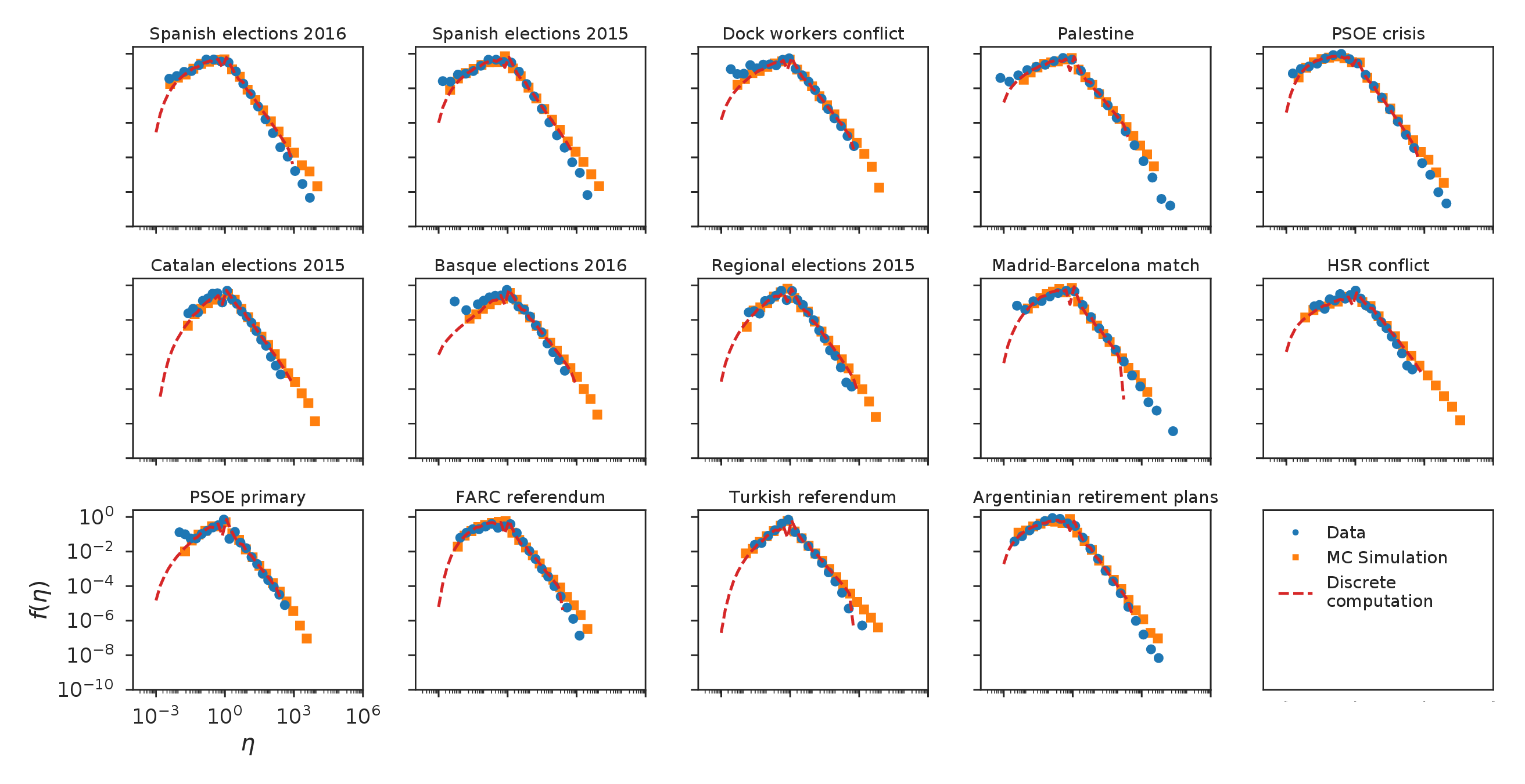}
\caption{
Efficiency distributions corresponding to the DiA model applied to the Twitter datasets. The plots show the empirical efficiency distribution (blue dots), the Monte-Carlo simulation (orange squares) and the discrete computation (dashed red line).}
\label{fig:twt_dip_eff}
\end{figure}

As discussed above, on the one hand, actors with low efficiency have a small impact on the system. Therefore, the InV model, where the interaction between system and individual is \emph{weak} (because the system’s response is not influenced by the individual’s activity) explains their distribution better. On the other hand, actors with high efficiency have a greater impact on the system, so it is the IdA model, where the interaction between individual and system is \emph{strong}, the one that explains their distribution better.
Since two models with different assumptions explain different intervals of the efficiency distribution and users with different efficiencies also present different levels of correlation between $A$ and $R$, we realize that the response of the system should depend on the features of each actor; that is, we should apply the DiA model.
In figure \ref{fig:twt_dip_eff} we present the results of applying the DiA model to the Twitter datasets by considering the number of followers as the feature that characterizes each actor. It can be appreciated that the DiA model behaves as expected and is able to reproduce both branches of the efficiency distribution for every Twitter dataset. 

Additionally, we have carried out a linear regression between the theoretical correlations $\rho_t$ induced by the DiA model and the empirical ones $\rho_e$ (see figure S4 of the SI), obtaining a significant correlation between them ($r^2=0.73$). The equation of the resulting line is $\rho_t = (1.03\pm0.18)\rho_e-(0.13\pm0.08)$. Notice that the slope is almost $1$ and is consistent with that value according to the standard error, indicating that the model reproduces the empirical correlations very faithfully.

The previous results confirm that the DiA model shows a very good agreement between theory and data both for the efficiency distribution and for the $A-R$ correlations. We conclude that in the case of Twitter it is necessary to consider the actors distinguishable to correctly describe the relationship between $A$ and $R$.

\section*{Discussion}


We have studied the relation between individual actions and the corresponding stimulated collective responses in social systems. To provide a deeper understanding of the systems, we have implemented three models that capture the essence of the dynamics between the two variables.
For each model we consider a different level of dependence between both magnitudes. We have used the distribution of the efficiency metric 
to relate activity $A$ and response $R$ and
evaluate the models in three social systems of different nature:  Twitter conversations, the scientific citations network and the Wikipedia collaboration environment. The theoretical efficiency distribution was computed using three methodologies: Monte-Carlo simulation, direct computation with discrete probability distributions and, for two of the models, we have derived analytical expressions.

In a previous work \cite{morales2014efficiency} it was found that the efficiency distribution was independent with respect to the $p(A)$ distribution, so it was hypothesized that $R$ was not affected by $A$. Following this line of reasoning, we developed the Independent Variables model to test if the universal structure of the efficiency distribution could be explained by the efficiency being the ratio of two independent random variables ($R$ and $A$) with heterogeneous distributions. We showed that to be true for the general shape but not when finer details are taken into account (some Twitter datasets present discrepancies between data and model in the right tails). Another relevant finding is that our analytical derivation of the InV model also explains the previous empirical findings regarding the independence of the efficiency distribution with respect to the activity distribution \cite{morales2014efficiency}. Besides, among the systems under study, the scientific citations data show the highest agreement with the InV model.



By studying the discrepancies in the right tails of the Twitter datasets for the InV model, we found that they could be explained by the presence of correlations that were unaccounted for by the InV model. Following these results, in the Identical Actors model, although all individuals are considered identical, the collective response of the system $R$ depends on the number of actions $A$ performed by an actor. When this model is tested on empirical data, we obtain excellent fits for the right tail of the efficiency distribution in the Twitter and Wikipedia datasets. Moreover, the IdA model reproduces the empirical correlations between $A$ and $R$ to a reasonable extent. However, the InV model showed a better agreement than the IdA model on the left tail of the efficiency distribution.

Since two models with different assumptions explained different intervals of the efficiency distribution, we hypothesized that the individual features of the actors must be considered to fully explain their efficiency, leading us to the development of the Distinguishable Actors model, in which the system's response depends on the characteristics of the individual that performs the action. 
We have applied this model to Twitter data by considering the number of followers of a user as a proxy of her ability to trigger a response in the system. Finally, we have shown that the DiA model presents a very good agreement with both tails of the empirical distribution of efficiency for the Twitter datasets and faithfully reproduces the empirical correlations between $R$ and $A$.

To conclude, we would like to stress that the adopted modeling approach, based on Ockham's razor principle, endows the models with a high explanatory power and provides fast and simple ways to compute them. It is also worth emphasizing the usefulness of the analytical expressions developed for the InV and IdA models, which enable immediate calculations. Although the developed formalism is general and domain-independent, the aforementioned properties also make them suitable to be used as null models for more elaborated and domain-specific approaches.

\section*{Methods}

\subsection*{Estimation of the PDF of efficiency from the discrete joint probability distribution of $A$ and $R$}

In order to obtain the efficiency distribution from the discrete computations of the IdA and DiA models we start from the cumulative probability distribution of efficiency:

\begin{equation}
P(H \leq \eta) = \sum_{A=1}^\infty \sum_{R=0}^{\lfloor A \eta \rfloor} p(R,A)
\label{eq:jp_2_eff_discr_CDF}
\end{equation}

Where $\lfloor \cdot \rfloor$ is the floor operator. Then, average probability densities for efficiency intervals $[\eta_a, \eta_b]$ can be computed as:

\begin{equation}
\bar{f}(\eta \in [\eta_a, \eta_b]) \approx \frac{P(H \leq \eta_b)-P(H \leq \eta_a)}{\eta_b - \eta_a}
\label{eq:jp_2_eff_discr_PDF}
\end{equation}

We adopt this approach to be able to compare our theoretical distributions with the empirical histograms.

\subsection*{Analytical approximation of the IdA model}

An analytical solution of the IdA model can be obtained if one is able to find an expression for the sum of $A$ independent random variables $\{r_1,r_2,\dots,r_A\}$ following the same distribution $p(r)$; that is, an expression for the $A-fold$ convolution of $p(r)$ with itself. 

Although finding an analytical expression for the sum of random variables is not always feasible, there exist approximations that work in some cases. We have modeled $p(r)$ as a power law and adopted an approximation proposed by Zaliapin et al. \cite{zaliapin2005approximating} to obtain the distribution of the sum of power-law distributed random variables. This approximation consists in replacing the sum by the maximum summand $R = \sum_{i=1}^A r_i \approx \max(\{r_i\}_{i=1}^A)$. In that case, if the PDF of $r$ is:

\begin{equation}
f_r(r) = (\gamma^{}_r -1)r_m^{\gamma^{}_r-1} r^{-\gamma^{}_r}
\end{equation}

The conditional cumulative distribution function of $R$ given $A$ can be computed as:

\begin{equation}
F(R|A) \propto e^{-( R^{1-\gamma^{}_r} A)}
\label{eq:F(R|A)_approx}
\end{equation}

Then, if the distribution of $A$ is modeled as a power law  with the form $p(A)\propto A^{\gamma^{}_A}$ and we assume that the minimum values of activity and response are $A_m =1$ and  $R_m = 1$ (which is what is observed for every dataset), the approximated joint probability density is:

\begin{equation}
\varphi(R,A) = \frac{(\gamma^{}_r-1)(\gamma^{}_A-1)}{1+(1-\gamma^{}_A)\Gamma(1-\gamma^{}_A,1)} A^{1-\gamma^{}_A}R^{-\gamma^{}_r}e^{-AR^{1-\gamma^{}_r}}
\end{equation}

Plugging this in  \eqref{eq:eff_anal_general}, we obtain the expression of \eqref{eq:idp_eff_pdf_anal_res}.

In principle, the expression of \eqref{eq:F(R|A)_approx} proposed in \cite{zaliapin2005approximating} is valid when $\gamma^{}_r < 2$. However, we have found that the expression of \eqref{eq:idp_eff_pdf_anal_res} also provides a good fit for the data when $\gamma^{}_r>2$, although the agreement quickly deteriorates when $\gamma^{}_r$ exceeds a value of around $2.5$.

 \subsection*{Formulation of the Independent Cascade model to compute the DiA model}

In the case of Twitter, the ability of the individuals to trigger a response depends mainly on their location on the follower network \cite{BORONDO201590}, meaning that it is a topocratic network \cite{Borondo2014}. In a previous work \cite{morales2014efficiency} it was also shown that the Independent Cascade (IC) model can be used to reproduce the efficiency distribution. Hence, in order to determine the  $p(r|\vec{s})$ distribution, we will rely on that application of the IC model to the Twitter follower network.


Taking into account that most information cascades in Twitter are shallow \cite{kwak2010twitter, bakshy2011everyone}, we have performed a first-neighbors approximation and focused on the response generated only in the first layer of diffusion. In that scenario, when a node becomes active (publishes a tweet), each follower can be {\it infected} with a probability $p_{inf}$, which we consider constant. If the active node has $F$ followers, this process can be formalized as $F$ Bernoulli trials (coin tosses) with success probability $p_{inf}$. Then, the response $r$ to this single action is a random variable with binomial distribution:

\begin{equation}
p(r|F) = B(r;F,p_{inf})
\end{equation}

The computation of the distribution of aggregate response $R$ is straight forward, as the sum of two binomial variables ($y=x_1+x_2$) with $n_1$ and $n_2$ trials and same success probability $p$ also follows a binomial distribution of the form $B(y;n_1+n_2,p)$. Therefore, we can consider that, if for a single action there are $F$ coin tosses, for $A$ actions we have $FA$ trials, obtaining the following distribution:

\begin{equation}
p(R|F,A) = B(R;FA,p_{inf})
\end{equation}
 
Notice that $p_{inf}$ can be considered as an effective infection probability that includes the effect of the higher layers. We have determined $p_{inf}$ by fitting the empirical single-action response distribution $p(r)$ to the corresponding theoretical distribution for this model through MLE:
 
 \begin{equation}
 p(r) = \sum_{F=0}^\infty B(r;F,p_{inf})p(F)
 \end{equation}
 
 Where $p(F)$ is the follower distribution. Once the $p_{inf}$ has been determined the joint probability distribution $p(R,A)$ can be computed as:
 
 \begin{equation}
 p(R,A) = p(A) \sum_{F=0}^\infty B(R;AF,p_{inf}) p(F)
 \label{eq:dip_jp_ra}
 \end{equation}
 
 And finally, the efficiency distribution can be determined with \eqref{eq:jp_2_eff_discr_CDF} and \eqref{eq:jp_2_eff_discr_PDF}. The numerical computations for the DiA model have been carried out with Twitter data considering the following cut-off values for the followers, the activity and the response: $F_{max} = 500000; A_{max} = 1000; R_{max} = 1000$.
 
 Although we have adopted this first-neighbors approximation mainly for simplicity and computational feasibility reasons, in the SI we present a formulation of the IC model that takes into account all the diffusion layers.

\bibliography{bibfile}

\section*{Acknowledgements}

This work has been supported by the Spanish Ministry of Science, Innovation and Universities (MICIU) under Contract No. PGC2018-093854-B-I00.

\section*{Author contributions statement}

SMG, JCL and RMB conceived and designed the research; SMG developed the formalism and analyzed the data; SMG, JCL and RMB interpreted the results and wrote the paper.

%

\clearpage
\appendix



\section*{Supplementary material}

# Supplementary Information for:
# Impact of individual actions on the collective response of social systems

S. Martin-Gutierrez, J. C. Losada and R. M. Benito

# Contents

# List of Figures

## List of Tables



In the first section of the Supplementary Information we present a detailed description of the datasets. In the second section, we derive the general formalism to compute the efficiency distribution from the activity and response distributions. In the third, the Independent Cascade model is fully formalized considering all the diffusion layers. In the fourth section, we explain the hybrid methodology adopted to fit several empirical probability distributions. In the fifth section we detail the computations carried out to measure the discrepancy between the InV model and the data. In the sixth section we discuss the empirical and theoretical correlations between $R$ and $A$. Finally, in the seventh section we present the fits of the data that have to be performed in order to compute the models.

# 1 Description of the datasets

We have worked with 29 datasets of three social systems of different nature: Twitter, the Wikipedia collaboration environment and the scientific citations network. Each dataset is thoroughly described below.

## 1.1 Twitter datasets

We have worked with 14 datasets extracted from different Twitter conversations. Each dataset has been retrieved by searching for tweets containing one of a given set of keywords during a specific period of time. In table S1 we present the keywords used to build each dataset as well as the number of users that took part in each conversation, the number of tweets they published and the time period considered in each case. From this information we have extracted the number of original tweets published by each user, which corresponds to the activity $A$, and the number of retweets obtained by each user, which is the response $R$. We have also computed the number of times each original tweet has been retweeted, which is the response of the system to single actions $r$.

| Dataset | Keywords | No. Tweets | No. Users | Time interval | Time span/days |
|---|---|---|---|---|---|
| Spanish elecions 2015 | 20D, 20D2015, #EleccionesGenerales2015 | 2779072 | 200159 | 2015/11/11 – 2016/05/09 | 180 |
| Spanish elections 2016 | 26J, 26J2016, #EleccionesGenerales2016, #Elecciones26J | 2814781 | 174110 | 2016/05/09 – 2016/11/13 | 188 |
| Dock workers conflict | estiba, estibadores | 218328 | 13489 | 2017/05/26 – 2017/07/14 | 49 |
| Palestine | Gaza, Ġazza, palestine, jerusalem, Yerusháláyim, Yerushalayim, yerussalem, quds, [same words in hebrew and arabic] | 3782389 | 372031 | 2017/12/20 – 2018/01/11 | 22 |
| PSOE crisis | psoe | 4766675 | 191260 | 2016/09/30 – 2016/11/04 | 35 |
| Catalan elections 2015 | 27s, 27s2015 | 90009 | 8130 | 2015/09/10 – 2015/09/15 | 5 |
| Basque elections 2016 | 25s | 97835 | 8535 | 2016/09/16 – 2016/09/27 | 11 |
| Spanish regional elections 2015 | 24m | 202865 | 20710 | 2015/05/21 – 2015/05/26 | 5 |
| Madrid-Barcelona match | madrid barsa, madrid barça, futbol madrid, fútbol madrid, futbol barcelona, fútbol barcelona, futbol barsa, fútbol barsa, fútbol barça, futbol barça, elclasico, elclásico, fcbarcelona, fcbarcelona_es, fcbarcelona_cat, realmadrid, clasico futbol, clásico futbol, clasico fútbol, clásico fútbol, clasico madrid, clásico madrid, clasico barsa, clasico barça, clásico barsa, clásico barça | 1394346 | 205991 | 2017/12/22 – 2017/12/25 | 3 |
| Murcia's HSR conflict | soterramientoya, soterramientomu, noalmuromurcia, alasvias, alasvías | 129617 | 4208 | 2017/10/06 – 2017/11/17 | 42 |
| PSOE primary | primarias psoe | 95874 | 13914 | 2017/05/20 – 2017/05/25 | 5 |
| FARC referendum | colombia acuerdo, colombia tratado, colombia farc, colombia referendum, referendum farc, tratado farc, acuerdo farc, colombia voto, colombia votar | 325050 | 40454 | 2016/10/07 – 2016/11/03 | 27 |
| Turkish referendum | Turkeyreferendum | 119250 | 13021 | 2017/04/15 – 2017/04/18 | 3 |
| Argentinian retirement plans | reforma argentina, disturbios argentina, pension argentina, pensión argentina, pensiones argentina, protesta argentina, protestar argentina, protestas argentina, violencia argentina, macri, mauriciomacri, cacerola argentina, cacerolas argentina | 4194063 | 265418 | 2017/12/21 – 2018/02/14 | 55 |

Table S1: Description of the Twitter datasets.

Here we provide some information on the contexts of the Twitter datasets:

- Spanish elections 2015: On 20 December 2015 the Spanish general elections were held [1, 2].

- Spanish elections 2016: On 26 June 2016, another general elections were held in Spain due to failure in the government formation process after the previous election [3].

- Dock workers conflict: A labor dispute around the liberalization of the stevedoring activity in Spanish docks [4].

- Palestine: This dataset tries to capture the online conversations regarding the conflict around Palestine.

- PSOE crisis: The Spanish Socialist's Workers Party (Partido Socialista Obrero Español) underwent a crisis that forced its leader dismissal on 28 September 2016 [5].

- Catalan elections 2015: Elections for the regional government of Catalonia held on 27 September 2015 [6].

- Basque elections 2016: Elections for the regional government of the Basque Country held on 25 September 2016 [7].

- Spanish regional elections 2015: Elections for the regional parliaments of thirteen out of the seventeen autonomous communities in Spain and local elections for all the municipalities held on 24 May 2015 [8].

- Madrid-Barcelona match: A football match held the 23 of December 2017 [9] between the two top-teams [10] of the Spanish football league: Real Madrid Club de Fútbol and Barcelona Fútbol Club.

- HSR conflict: Protests in the city of Murcia, Spain, around the building of a wall for the high speed rail (HSR) that would divide the city [11].

- PSOE primary: The election held by the Spanish Socialist's Workers Party to choose new leadership after the crisis of 2016 [12].

- FARC referendum: A referendum to ratify the final agreement on the termination of the Colombian conflict between the Colombian government and the FARC guerillas was held on 2 October 2016. It failed with 50.2% voting against it and 49.8% voting in favor [13].

- Turkish referendum: A constitutional referendum was held throughout Turkey on 16 April 2017 on whether to approve 18 proposed amendments to the Turkish constitution that would replace the existing parliamentary system of government by an executive presidency [14].

- Argentinian retirement plans: A reform of the public retirement plans system in Argentina passed by the government on 19 December 2017 triggered violent protests [15].

## 1.2 Scientific citations datasets

We have queried the Web of Science [16] to extract all the scientific papers published in several countries as well as their authors and the citation data for both papers and authors. The data spans from 2008 to 2017. Although the authors are associated to a given country, the citations they receive come from any author in any country. In table S2 we show the number of papers and authors for each country. In these datasets, the number of papers published by a given author is her activity $A$ and the total number of citations she has obtained, the response $R$. The single-action response $r$ in this case corresponds to the number of citations to a given paper.

| Dataset | No. Publications | No. Authors |
|---|---|---|
| Argentina | 89070 | 56095 |
| Brazil | 406493 | 211829 |
| Chile | 68375 | 53318 |
| Colombia | 41106 | 45983 |
| Cuba | 10570 | 8723 |
| Finland | 118223 | 82957 |
| Greece | 106241 | 71089 |
| Iceland | 9623 | 16725 |
| Israel | 133219 | 75765 |
| New Zealand | 86875 | 54900 |
| Poland | 233673 | 122556 |
| Portugal | 122848 | 85512 |
| Spain | 524770 | 261740 |
| Venezuela | 12488 | 8921 |

Table S2: Description of the datasets of scientific citations.

## 1.3 Wikipedia dataset

The Wikipedia dataset has been retrieved from the Stanford Large Network Dataset Collection [17]. It contains data about the editions made by users on the Wikipedia pages from the beginning of the Wikipedia project on 15 January 2001 to January 2008. In this case, we have only one dataset, which corresponds to the English Wikipedia, that includes 250 million registered editions by 2.3 million users. In Wikipedia users can edit regular article pages, the talk pages associated to each article to coordinate the redaction and also personal user pages, that are intended as a way of exchanging direct messages between users. In this system, we have considered the number of editions performed by a user as her activity $A$ and the number of editions made by other users in her personal page as the response $R$. In this case, there is no direct way to define the response of the system to single actions as in the other two studied systems. Hence, we have defined $r$ as the number of editions made on the personal pages of users that have performed only one edition themselves; that is, users whose activity $A$ is 1.

## 2 Computing the efficiency distribution with random variable algebra

For the sake of clarity, only for this section we will adopt the usual convention in statistics of writing the name of a random variable with an uppercase letter ($X$) and a specific value of that variable as the corresponding lowercase letter ($x$). Therefore, when we write $r$ in this section, we are referring to a specific value of the response random variable $R$, not to the partial response to a single action by a given actor.

The distribution of efficiency can be obtained from the joint probability distribution of $A$ and $R$ with random variable algebra [18, 19]. Let $A$ and $R$ be two random variables with support $a \in [a_m, \infty)$ and $r \in [r_m, \infty)$ for some $a_m, r_m > 0$. Efficiency ($H$ - uppercase $\eta$) is another random variable defined as $H = \frac{R}{A}$. Let $\eta$ be a particular value of $H$. The cumulative probability that $H \leq \eta$ can be written as:

$$p(H \leq \eta) = p(\frac{R}{A} \leq \eta) = p(R \leq \eta A) = p(A \geq \frac{R}{\eta}) \tag{S1}$$

Where we have replaced $H$ by $R/A$. Let us now make the approximation that $A$ and $R$ are continuous variables with joint probability distribution $\varphi_{R,A}(r, a)$ (which we will call $\varphi(r, a)$ to simplify notation). We

can compute the cumulative distribution function (CDF) of $H$; that is, $F_H(\eta) = p(H \leq \eta)$, by integrating $\varphi(r,a)$ in the region where Eq. (S1) is fulfilled:

$$F_H(\eta) = \begin{cases} \int_{r_m}^{\infty} \int_{r/\eta}^{\infty} \varphi(r,a) da dr & \text{if} \quad \eta \leq \frac{r_m}{a_m} \\ \int_{a_m}^{\infty} \int_{r_m}^{\eta a} \varphi(r,a) dr da & \text{if} \quad \eta > \frac{r_m}{a_m} \end{cases} \quad (S2)$$

In figure S1 we show a diagram explaining the regions of integration. Notice that in the two-dimensional $(r, a)$ space, the subspaces of constant efficiency are straight lines that pass through $(0, 0)$ because, for a specific efficiency $\eta_* \neq 0$, we can write $a = \frac{1}{\eta_*} r$. Therefore, lines with lower $\eta_*$ have higher slope and vice versa. The probability that the efficiency takes a value smaller than a given $\eta_*$; that is, $p(H \leq \eta_*)$, can be found by integrating the probability density $\varphi(r, a)$ to the *left* of the $a = \frac{1}{\eta_*} r$ line (integrate for all the efficiencies $\eta \leq \eta_*$, so lines with higher slope than $1/\eta_*$). Now, because the minimum values $a_m, r_m$ are larger than 0, the regions of integration must be defined differently to each side of the line with $\eta_o = \frac{r_m}{a_m}$ (thick dashed line of the diagram). The green area is an example of area of integration associated to a particular efficiency value $\eta_1 < \eta_o$. There, the integral in variable $a$ runs from the line defined by $a = \frac{1}{\eta_1} r$ to infinity and in variable $r$ from $r_m$ to infinity. When we integrate for a efficiency value $\eta_2 > \eta_o$, the area of integration, an example of which is shown in orange, is defined by taking variable $r$ running from $r_m$ to the line $r = \eta_2 a$ and variable $a$ from $a_m$ to infinity. If we used integration regions of the *green kind* in the area where $\eta > \eta_o$, we would incorrectly include the red area in the integral, while if we used integration regions of the *orange kind* in the area where $\eta < \eta_o$, we would incorrectly include the yellow area in the integral. Therefore, we must compute $F_H(\eta)$ differently to each side of $\eta_o = \frac{r_m}{a_m}$ to obtain the right result. The two branches of Eq. (S2) correspond to the two tails of the efficiency distribution.

Once we have the cumulative distribution function $F_H(\eta)$, the Leibniz integral rule can be used to get $f_H(\eta) = \frac{dF_H(\eta)}{d\eta}$. For $\eta \leq \frac{r_m}{a_m}$:

$$f_H(\eta) = \frac{d}{d\eta} \left[ \int_{r_m}^{\infty} \int_{r/\eta}^{\infty} \varphi(r,a) da dr \right] = \int_{r_m}^{\infty} \varphi(r, \frac{r}{\eta}) \left( \frac{r}{\eta^2} \right) dr$$

Change of variables $\quad r = \eta a \quad \Rightarrow \quad f_H(\eta) = \int_{r_m/\eta}^{\infty} \varphi(\eta a, a) a da$ (S3)

And for $\eta > \frac{r_m}{a_m}$:

$$f_H(\eta) = \frac{d}{d\eta} \left[ \int_{a_m}^{\infty} \int_{r_m}^{\eta a} \varphi(r,a) dr da \right] = \int_{a_m}^{\infty} \varphi(\eta a, a) a da \quad (S4)$$

We use Eq. (S3) and Eq. (S4) to get the analytical expressions of the InV and IdA models. They are rewritten in Eq. (2) of the main document dropping the uppercase/lowercase convention adopted only for this section, so $R$ is used instead of $r$ and $A$ is used instead of $a$.

## 3  Full Independent Cascade model formalism

In this section, we develop the formalization of the Independent Cascade model adopted to compute the DiA model for the Twitter data. Here we start by considering all the diffusion layers and then carry out the first-neighbors approximation used for the computations presented in the main document.

The IC model consists in a diffusion mechanism [20] that can be outlined as follows. Assume that we have a set of agents (nodes) that are connected with a set of links forming a static directed network (the links do not change through time). Choose one node randomly to be activated. The active node tries to activate each of her neighbors one by one. There is only one attempt per neighbor and all the attempts are independent. The success probability for each activation attempt depends on the link that joins the source

node and the destiny node. The successfully *infected* nodes become active and try to activate their neighbors. This process is iterated until the diffusion cascade dies out.

Now, drawing on the work by Morales et al. [21], let us assume that the probability of a given node to activate her neighbors only depends on her distance to the source of the cascade. If she is the source, the *infection* probability will be $p_{inf}^1$. If she is $l$ layers away from the source, the infection probability will be $p_{inf}^l$. The number of *neighbors* of a node is her out-degree $k_{out}^l$ (considering the direction of the links as the direction of propagation of influence), which we will call $k^l$ in order to simplify the notation.

Taking the above into account, let us try to find the response distribution conditioned on the activity. Notice that if we look at a given active node trying to *infect* her neighbors we can model this process with a binomial variable: In the first layer, the node has $k^1$ Bernoulli trials to activate a neighbor and each trial has a success probability $p_{inf}^1$. The number of successfully activated neighbors in that layer corresponds to the response $r^1$, which can take values between 0 and $k^1$. The response distribution for the first layer of the diffusion cascade would be

$$p(r^1|k^1) = B(r^1; k^1, p_{inf}^1) \tag{S5}$$

With $B(x; n, p)$ a distribution of a binomial variable $x$ with $n$ Bernoulli trials and $p$ success probability.

An individual that performs $A$ actions and has $k^1$ neighbors will trigger an aggregated response $R^1 = \sum_{i=1}^{A} r_i^1$ in her immediate neighbors (the first layer). The distribution of the aggregated response can be computed straight forward, as the sum of two binomial variables ($y = x_1 + x_2$) with $n_1$ and $n_2$ trials and same success probability $p$ also follows a binomial distribution of the form $B(y; n_1 + n_2, p)$. The probability distribution for $R^1$ given a specific activity $A$, an out-degree $k^1$ and an infection probability $p_{inf}^1$ is then:

$$p(R^1|A, k^1) = B(R^1; Ak^1, p_{inf}^1) \tag{S6}$$

Our next approximation is that a response of $R^1$ implies that $R^1$ nodes were activated (the reality is that there were $R^1$ activations of a total of $k^1$ nodes, some of them possibly activated more than once). The $R^1$ active nodes have $k_1^2, k_2^2, \ldots, k_{R^1}^2$ neighbors (the superscript 2 indicates that they are in the second diffusion layer). Each of the active nodes tries to *infect* her neighbors with the same infection probability $p_{inf}^2$. Hence, there are a total of $k_T^2(R^1) = \sum_{i=1}^{R^1} k_i^2$ Bernoulli trials with success probability $p_{inf}^2$, resulting in the following distribution of aggregated response for the second diffusion layer:

$$p(R^2|R^1, k_T^2(R^1)) = B(R^2; k_T^2(R^1), p_{inf}^2) \tag{S7}$$

We can marginalize this with respect to $R^1$ and $k_T^2(R^1)$ in order to get the distribution of $R^2$, but first we need to obtain the distribution of $k_T^2(R^1)$, which can be computed as the $R^1 - fold$ convolution of the empirical $p(k)$ distribution with itself:

$$p(k_T^2|R^1) = p^{*R^1}(k) \tag{S8}$$

In order to be able to do this, we have to assume that $p(k^{l+1}|k^l) = p(k)$; that is, that there exists no degree correlations. This is a reasonable approximation for Twitter because it has been observed that the degree correlations are small ($|\rho| < 0.2$) [22, 2, 23], a property that is also present in other online social networks [24].

The probability distribution of $R^2$ is finally:

$$p(R^2|A,k^1) = \sum_{k_T^2=0}^{\infty} \sum_{R^1=0}^{\infty} p(R^2|R^1,k_T^2(R^1))p(k_T^2|R^1)p(R^1|A,k^1)$$
$$= \sum_{k_T^2=0}^{\infty} \sum_{R^1=0}^{\infty} B(R^2;k_T^2(R^1),p_{inf}^2)p^{*R^1}(k)B(R^1;Ak^1,p_{inf}^1) \quad (S9)$$

This process can be iterated to get the response distribution for layer $l$ as a function of the one obtained in layer $l-1$:

$$p(R^l|A,k^1) =$$
$$= \sum_{k_T^l=0}^{\infty} \sum_{R^{l-1}=0}^{\infty} p(R^l|R^{l-1},k_T^l(R^{l-1}))p(k_T^l|R^{l-1})p(R^{l-1}|A,k^1)$$
$$= \sum_{k_T^l=0}^{\infty} \sum_{R^{l-1}=0}^{\infty} B(R^l;k_T^l(R^{l-1}))p^{*R^{l-1}}(k)p(R^{l-1}|A,k^1) \quad (S10)$$

The total aggregated response triggered in all the layers would be $R = \sum_{l=1}^{\infty} R^l$. Hence, the distribution of the response given an activity $A$ and a number of neighbors $k^1$ is the convolution of the distributions of response for each layer:

$$p(R|A,k^1) = p(R^1|A,k^1) * p(R^2|A,k^1) * \cdots \quad (S11)$$

Finally, the joint probability distribution for $R$ and $A$ can be obtained as:

$$p(R,A) = p(A) \sum_{k^1=0}^{\infty} p(R|A,k^1)p(k^1) \quad (S12)$$

Notice that $p(k^1) = p(k)$. In this model, the infection probabilities $\{p_{inf}^l\}_{l=1}^{\infty}$ are free parameters that must be determined in some way, be it by direct computation if the data is available, by Maximum Likelihood Estimation or any other fitting technique.

The computations in this model comprehend an infinite number of diffusion layers, making it numerically intractable. We have chosen, as a first order approximation, to consider only the response of the first layer. The rationale behind this approximation is that it has been observed that most information cascades in Twitter are shallow [25, 26]. Moreover, the computations involved for the second and farther layers are very demanding in terms of memory and computation time, and the results that we obtain for the first-layer approximation are already in good agreement with the data.

Since we work with the follower network of Twitter, the out-degree distribution $p(k)$ corresponds to the follower distribution $p(F)$. Hence, for our first-layer approximation to the independent cascade model, the distribution of response $R$ given the activity $A$ and a certain number of followers $F$ is:

$$p(R|A,F) = B(R;AF,p_{inf}) \quad (S13)$$

Where we use an effective infection probability $p_{inf}$ that includes the effect of the higher layers. We have determined $p_{inf}$ by fitting the empirical distribution of response to single actions $p(r)$ to the corresponding theoretical distribution for this model with MLE:

$$p(r) = \sum_{F=0}^{\infty} B(r;F,p_{inf})p(F) \quad (S14)$$

10

Where $p(F)$ is the empirical follower distribution. Once the $p_{inf}$ has been determined it is straight forward to compute the joint probability distribution $p(R, A)$:

$$p(R, A) = p(A) \sum_{F=0}^{\infty} B(R; AF, p_{inf}) p(F) \tag{S15}$$

# 4 Hybrid fitting methodology for the $p(r)$ and $p(F)$ distributions

The hybrid methodology has been developed to fit several distributions in order to obtain the most faithful reproduction of the data, such that the errors committed when performing numerical computations are minimized.

Since the empirical distributions of discrete variables $k$ analyzed in this work present a heavy tail (see SI), the highest probabilities are associated to low values of $k$, where the empirical distribution is very well defined. The smaller probabilities, which correspond to high values of $k$, show a more noisy behavior due to a lack of data. Also, beyond a certain threshold there are no more data points.

Taking this into account, the empirical distribution has been fit by taking the raw empirical values $p_k^{emp}$ of the probabilities for low $k$ and then fitting the tail of the distribution to a theoretical one with probabilities $p_k^{fit}$. In particular, the lognormal and the truncated power law distributions have been employed. The tail is defined as those values of $k$ which are larger than a given threshold $q$. The resulting hybrid distribution has the following form:

$$p(k) = \begin{cases} p_k^{emp} & if \quad k \leq q \\ C p_k^{fit} & if \quad k > q \end{cases} \tag{S16}$$

Where the normalization constant $C$ can be obtained by properly normalizing $p(k)$:

$$C = \sum_{k=q}^{\infty} p_k^{emp} \tag{S17}$$

With this methodology a very faithful representation of $p(k)$ is obtained and we are able to compute the probabilities for values of $k$ that are missing in the data; in particular, very high and unlikely values.

The distribution of response to single actions has been modeled with this methodology using a lognormal and the follower distribution has been fit employing a truncated power law for the tail. The resulting fits are presented below.

# 5 Measurement of the discrepancy between the InV model and the data and its relationship with empirical $A - R$ correlations

The discrepancy between the InV model and the data can be characterized by the disagreement in the right tails of the efficiency distributions. This disagreement has been determined by modeling the right tail of the empirical efficiency distribution as a power law ($p(\eta) \propto \eta^{-\gamma_{emp}}$) and computing the difference between the empirical $\gamma_{emp}$ exponent and the $\gamma_{InV}$ exponent according to the InV model. For this purpose, the response distribution was modeled as a power law with exponent $\gamma_R$. As we discussed in the main document, in that case $\gamma_{InV} = \gamma_R$. Consequently, the measure for the deviation between the InV model and the empirical results can be defined as:

$$\Delta_{InV} = \gamma_{emp} - \gamma_{InV} = \gamma_{emp} - \gamma_R \tag{S18}$$

In figure S2 we plot the error associated to the InV model $\Delta_{InV}$ against the empirical Spearman's correlation between $A$ and $R$ ($\rho_e$) and each point corresponds to a different Twitter conversation. It can be appreciated that $\Delta_{InV}$ grows monotonously with $\rho_e$.

# 6 Theoretical and empirical $A - R$ correlations

To show that the IdA and DiA models reproduce the empirical $A - R$ correlations we generate 300 MC simulations of the models for each Twitter dataset and for every simulation we compute the corresponding $A - R$ correlation ($\rho_t$). Then we compute the mean and the standard deviation of the 300 MC simulations to get a confidence interval for the theoretical correlations.

In figure S3 we plot the theoretical Spearman's correlation according to the IdA model ($\rho_t$) against the empirical one ($\rho_e$) for the Twitter datasets and carried out a linear regression between both magnitudes. As it can be appreciated, there is a significant correlation between the empirical and the theoretical correlations. Moreover, the slope is close to 1 and the value of the intercept is close to 0. This confirms that the IdA model reproduces the correlations of the real data to a reasonable extent.

The correlations that the DiA model induces between $A$ and $R$ ($\rho_t$) have been represented against the empirical correlations ($\rho_e$) in figure S4. The linear regression that is included in the aforementioned figure shows that, analogously to the IdA model, there exists a significant correlation between empirical and theoretical results. Additionally, it can be appreciated that in the case of the DiA model the slope is closer to 1 and, more importantly, consistent with that value according to the standard error. Hence, the DiA model improves the results of the IdA model and reproduces the correlations between $A$ and $R$ more faithfully.

# 7 Intermediate computations and fits

In this section we detail all the intermediate computations that need to be performed in order to obtain the efficiency distribution for each one of the three proposed models: the independent variables (InV) model, the identical actors (IdA) model and the distinguishable actors (DiA) model. In particular, we present fits of the data to heavy-tailed distributions performed with the powerlaw python package [27].

## 7.1 Independent variables model

In the InV model we fit the probability density functions (PDFs) of activity $A$ and response $R$ to two different statistical distributions: the power law (PL) and the exponentially truncated power law (TPL). The former is used to characterize the deviation of this model from the empirical results and the latter to reproduce the empirical distribution of efficiency. In this section, we present the resulting fits of the data to those models.

The expression employed for the power law distribution is the following:

$$f(x) = (\gamma - 1)x_{min}^{\gamma-1}x^{-\gamma} \tag{S19}$$

Where $x_{min}$ corresponds to the minimum value of $x$. The exponentially truncated power law distribution can be written as:

$$f(x) = \frac{\lambda^{1-\alpha}}{\Gamma(1 - \alpha, \lambda x_{min})}x^{-\alpha}e^{-\lambda x} \tag{S20}$$

### 7.1.1 Scientific citations

The efficiency distribution according to the InV model in the case of the scientific citations datasets was computed by adopting the truncated power law models for the $f_R(R)$ and $f_A(A)$ PDFs. However, as the empirical distributions present a very noisy behavior towards the tail, instead of using the whole datasets, the data selected to perform the fits corresponded to $A \in [1, 30]$ and $R \in [2, 300]$. The resulting fits are shown in green in figures S5 and S6, where the raw data are represented as blue dots and the data used to perform the fits as orange squares. Notice that, although the fits were not performed using the full range of the data, the distributions of efficiency shown in the main document were computed without filtering out any data. The parameters $\alpha$ and $\lambda$ of Eq. (S20) computed for each fit are presented in table S3.

### 7.1.2 Twitter

The distributions of activity and response for the Twitter datasets were also modeled with TPL. Additionally, the distribution of response and the right tail of the distribution of efficiency for Twitter were also fit to a regular PL in order to determine the departure of the InV model with respect to the data. The fits that have been obtained are presented in figures S7, S8 and S9. The parameters of Eq. (S19) and Eq. (S20) for the response distributions as well as the parameters of Eq. (S20) for the activity distribution are presented in table S4.

### 7.1.3 Wikipedia

In the case of the Wikipedia dataset, the distributions of $A$ and $R$ have been modeled with TPL. The resulting fits are presented in figures S10 and S11 and the parameters of Eq. (S20) in table S5.

## 7.2 Identical actors model

The IdA model requires the distribution of response to single actions $p(r)$ and the distribution of activity $p(A)$ as InVuts. In the main document three different methodologies are proposed to solve this model: Monte-Carlo simulation, direct numerical computation (Eq. (10), Eq. (17) and Eq. (18) of the main document) and an analytical approximation (Eq. (11) of the main document). In the first two methodologies a hybrid approach (explained above) is adopted to fit the empirical distributions of $r$ for some of the systems (for others, a regular fit is performed). Here, we present the resulting fits for every system.

In the hybrid methodology, the tail of the $p(r)$ distribution is modeled as a lognormal with the following expression:

$$f(x) = \sqrt{\frac{2}{\pi\sigma^2}} \left[ \text{erfc}\left(\frac{\log x_{min} - \mu}{\sqrt{2}\sigma}\right) \right] \frac{1}{x} e^{-\frac{(\log x - \mu)^2}{2\sigma^2}} \tag{S21}$$

With respect to the analytical approximation of the model (Eq. (11) of main document), both $r$ and $A$ are modeled as continuous random variables distributed as power laws. These fits are also presented in this section.

### 7.2.1 Scientific citations

The distribution of activity $p(A)$ of the scientific citations network for the numerical computation of the IdA model was modeled as a TPL and the resulting fit was already shown in figure S5. In order to compute the analytical approximation of the efficiency distribution, $p(A)$ was also fit to a power law. The resulting fits are shown in figure S12 and the corresponding parameters of Eq. (S19) in table S3.

In order to compute the IdA model numerically, The $p(r)$ on the other hand was fit to a lognormal distribution using the hybrid methodology for the interval $r \in [1, \infty)$ and maintaining the empirical probability for $r = 0$. The results are shown in figure S13 and the parameters of the fit are presented in table S3. For the analytical approximation of the IdA model, we have modeled $p(r)$ as a PL. We have fit the data of $r$ to a PL by computing the optimal $r_{min}$ that optimizes the Kolmogorov-Smirnov distance [28] with the methodology implemented in [27]. The results are shown in figure S14 and the parameters of the fits presented in table S3. As can be seen, whereas the hybrid distributions of figure S13 show a very good agreement with the data, the PL fits of figure S14 are only good for the tail. The lower values of $r$ present a large divergence between the data and the fitting line. This causes the analytical approximation of the IdA model to perform poorly in the case of the scientific citations datasets, as will be discussed in section 7.2.4.

### 7.2.2 Twitter

The activity distribution of the Twitter datasets for the numerical computation of the IdA model was modeled as a TPL, so the resulting fit was already shown in figure S7. In order to compute the analytical approximation of the efficiency distribution, the $p(A)$ was also fit to a power law. The resulting fits are also shown in figure S7 and the corresponding parameters of Eq. (S19) in table S4.

The $p(r)$ on the other hand was fit using the hybrid methodology to a lognormal distribution for $r \in [1, \infty)$ maintaining the empirical probability for $r = 0$ in order to compute the IdA model numerically. The results are shown in figure S15 and the parameters of the fit are presented in table S4. For the analytical approximation of the IdA model, we have modeled $p(r)$ as a PL. We have fit the data of $r$ to a PL following the same methodology as for the scientific citations datasets. The results are shown in figure S16 and the parameters of the fits presented in table S4.

### 7.2.3 Wikipedia

The distribution of activity in the case of Wikipedia was fit to a TPL to compute the IdA model. The result was shown in figure S10. The analytical approximation of the model was computed with a PL fit of the activity distribution, which is also presented in figure S10. The parameters of Eq. (S19) can be consulted in table S5.

The $p(r)$ on the other hand was fit using the hybrid methodology to a PL distribution for $r \in [4, \infty)$ maintaining the empirical probability for $r \in [0, 4)$ in order to compute the IdA model numerically. The results are shown in figure S17 and the parameters of the fit are presented in table S5. For the analytical approximation of the IdA model, we have modeled $p(r)$ as a PL. We have fit the data of $r$ to a PL following the same methodology as for the scientific citations and Twitter datasets. The results are shown in figure S18 and the parameters of the fits presented in table S5.

### 7.2.4 The error associated to the PL fit of $p(r)$ affects the quality of the IdA model's analytical approximation

In the main document it is observed that the analytical approximation of the IdA model shows different levels of agreement with the Monte-Carlo simulations and the numerical computations. Here we show that the main cause for this deviation is the error associated to the power law fit of the distribution of response to single actions $p(r)$.

We have focused on the differences in the right tail of the efficiency distribution between the numerical and analytical computations. Given the heterogeneous nature of this distribution, we have measured these differences as the mean squared error of the logarithm of the probabilities. Therefore, if $f_{num}(\eta_i)$ is the average probability density associated to interval $i$ of the histogram according to the numerical computation

14

and $f_{appr}(\eta_i)$ is the average probability density associated to the same interval according to the analytical derivation, the mean squared error $\epsilon_\eta$ would be:

$$\epsilon_\eta = \sqrt{\frac{\sum_{i=1}^n (\log[f_{appr}(\eta_i)] - \log[f_{num}(\eta_i)])^2}{n}} \quad (S22)$$

Where $n$ is the number of bins used to compute the histogram. We have measured the error committed when $p(r)$ is fit to a power law in an analogous way. Let $p_r^{emp}$ the empirical probability associated to value $r$ of the response to single actions and $p_r^{fit}$ the probability associated to the same value according to the power law fit. The mean squared error $\epsilon_r$ associated to this fit would be:

$$\epsilon_r = \sqrt{\frac{\sum_r (\log[p_r^{fit}] - \log[p_r^{emp}])^2}{n}} \quad (S23)$$

In this case the sum runs over all the possible values of $r$ with non-zero empirical probabilities.

In figure S19 the relationship between $\epsilon_r$ and $\epsilon_\eta$ can be clearly appreciated. It is worth pointing out that this relationship seems to follow the same tendency for all the systems under study. Figure S19 also shows that the scientific citations system presents the worse agreements between the analytical approximation and numerical computation of the IdA model due to the lower quality of the PL fits of the $p(r)$ distribution.

## 7.3 Distinguishable actors model

The DiA model is defined by the conditional distribution of response for single actions $r$ given the feature vector $\vec{s}$ of the individual: $p(r|\vec{s})$. Twitter was the only system analyzed under this model because of the inherent complication of defining and quantifying the relevant characteristics of an individual that determine her ability to trigger a response in an arbitrary setting. In the case of Twitter, the response only depends on the number of followers $F$ of the individual and on an infection probability $p_{inf}$ that is considered constant for every action and every individual:

$$p(r|\vec{s}) = p(r|F) = B(r; F, p_{inf}) \quad (S24)$$

Where $B(k; n, p)$ is a binomial distribution of a variable $k$ with $n$ Bernoulli trials and $p$ success probability. The distribution of efficiency is computed in this case with Eq. (16), Eq. (17) and Eq. (18) of the main document.

Consequently, in order to solve the DiA model for Twitter it is necessary to compute the value of $p_{inf}$, which is accomplished by performing a Maximum Likelihood Estimation (MLE) of $p(r)$, and fit $p(F)$, which is carried out with the hybrid methodology presented in the Materials and Methods section of the main document. In this section we present the results for the MLE estimation of $p_{inf}$ and the fits of the follower distribution $p(F)$.

The effective $p_{inf}$ of a given dataset can be computed by maximizing the following log-likelihood function:

$$\mathcal{L} = \sum_i \log\left[\sum_F p(F) B(r_i^{obs}; F, p_{inf})\right] \quad (S25)$$

Where $r_i^{obs}$ are observed values of response to single actions. However, given the enormous quantity of data, the optimization process can be very slow. In order to perform the computations faster, samples of the $r$ data for each dataset were employed instead of the whole dataset. Specifically, we took 10 different samples of 1000 $r_i^{obs}$ each and computed the average $p_{inf}$ for each dataset, obtaining also a standard deviation. The results are shown in table S6.

15

The hybrid distribution of $p(F)$, which has been obtained by modeling the tail of the distribution with a PL, is shown in figure S20 and the corresponding parameters in table S4.

Table S3: Parameters of each fit performed with the scientific citations datasets.

| Dataset | A | | | R | | | η | r | | | | PL |
|---|---|---|---|---|---|---|---|---|---|---|---|---|
| | TPL | | PL | TPL | | PL | PL | Hybrid lognormal | | | | |
| | $\alpha$ | $\lambda$ | $\gamma$ | $\alpha$ | $\lambda$ | $\gamma$ | $\gamma$ | $q$ | $\mu$ | $\sigma$ | $x_{min}$ | $\gamma$ |
| Argentina | 2,273 | 8,237E-02 | 2,745 | 1,106 | 1,174E-02 | 1,564 | 1,696 | 1 | 2,085 | 1,174 | 31 | 2,676 |
| Brazil | 1,777 | 1,004E-01 | 2,442 | 1,443 | 5,039E-03 | 1,663 | 1,824 | 1 | 1,905 | 1,168 | 37 | 2,896 |
| Chile | 2,058 | 8,609E-02 | 2,592 | 1,062 | 9,894E-03 | 1,524 | 1,696 | 1 | 2,087 | 1,282 | 46 | 2,732 |
| Colombia | 2,324 | 2,839E-02 | 2,540 | 1,000 | 6,063E-03 | 1,438 | 1,702 | 1 | 1,851 | 1,373 | 29 | 2,541 |
| Cuba | 2,757 | 4,929E-03 | 2,793 | 1,000 | 5,689E-03 | 1,401 | 1,855 | 1 | 1,841 | 1,286 | 37 | 2,698 |
| Finland | 1,890 | 7,885E-02 | 2,438 | 1,050 | 1,388E-02 | 1,560 | 1,723 | 1 | 2,486 | 1,184 | 63 | 2,679 |
| Greece | 2,134 | 6,302E-02 | 2,554 | 1,127 | 1,046E-02 | 1,560 | 1,841 | 1 | 2,249 | 1,219 | 39 | 2,705 |
| Iceland | 2,239 | 1,058E-01 | 2,808 | 1,000 | 8,178E-03 | 1,413 | 1,489 | 1 | 2,566 | 1,363 | 26 | 2,130 |
| Israel | 2,192 | 5,861E-02 | 2,581 | 1,137 | 1,152E-02 | 1,576 | 1,733 | 1 | 2,319 | 1,271 | 95 | 2,760 |
| New Zealand | 2,121 | 5,215E-02 | 2,494 | 1,121 | 9,706E-03 | 1,548 | 1,685 | 1 | 2,341 | 1,199 | 55 | 2,693 |
| Poland | 1,694 | 8,122E-02 | 2,310 | 1,250 | 6,862E-03 | 1,577 | 1,835 | 1 | 1,840 | 1,241 | 30 | 2,617 |
| Portugal | 2,055 | 5,369E-02 | 2,450 | 1,140 | 1,053E-02 | 1,567 | 1,745 | 1 | 2,324 | 1,181 | 51 | 2,890 |
| Spain | 1,687 | 1,056E-01 | 2,401 | 1,332 | 9,494E-03 | 1,657 | 1,911 | 1 | 2,349 | 1,211 | 87 | 2,837 |
| Venezuela | 1,008 | 4,269E-01 | 2,873 | 1,183 | 1,156E-02 | 1,599 | 1,585 | 1 | 1,679 | 1,364 | 38 | 2,542 |

Table S4: Parameters of each fit performed with the Twitter datasets.

| Dataset | A | | | R | | | η | r | | | | PL | | F | |
|---|---|---|---|---|---|---|---|---|---|---|---|---|---|---|---|
| | TPL | | PL | TPL | | PL | PL | Hybrid lognormal | | | PL | | | Hybrid TPL | |
| | $\alpha$ | $\lambda$ | $\gamma$ | $\alpha$ | $\lambda$ | $\gamma$ | $\gamma$ | $q$ | $\mu$ | $\sigma$ | $x_{min}$ | $\gamma$ | $q$ | $\alpha$ | $\lambda$ |
| Spanish elections 2016 | 1,864 | 5,407E-03 | 1,935 | 1,471 | 5,478E-04 | 1,535 | 1,937 | 1 | -0,136 | 1,889 | 125 | 2,509 | 500 | 1,806 | 1,067E-08 |
| Spanish elections 2015 | 1,903 | 3,891E-03 | 1,956 | 1,476 | 7,432E-04 | 1,547 | 1,965 | 1 | -0,193 | 1,846 | 60 | 2,465 | 500 | 1,813 | 1,100E-08 |
| Dock workers conflict | 2,066 | 2,352E-03 | 2,092 | 1,477 | 4,368E-04 | 1,535 | 1,736 | 1 | -1,665 | 2,607 | 41 | 2,119 | 500 | 1,625 | 3,462E-07 |
| Palestine | 2,171 | 1,614E-03 | 2,187 | 1,576 | 2,086E-04 | 1,605 | 1,725 | 1 | -3,197 | 2,864 | 41 | 2,101 | 500 | 1,610 | 9,378E-08 |
| PSOE crisis | 1,597 | 5,649E-03 | 1,719 | 1,470 | 2,550E-04 | 1,519 | 1,850 | 1 | -1,877 | 2,434 | 222 | 2,549 | 500 | 1,817 | 1,368E-07 |
| Catalan elections 2015 | 2,107 | 1,009E-02 | 2,179 | 1,387 | 2,506E-04 | 1,526 | 1,854 | 1 | 0,035 | 1,806 | 10 | 2,138 | 500 | 1,789 | 5,998E-07 |
| Basque elections 2016 | 2,302 | 9,530E-07 | 2,302 | 1,405 | 1,544E-03 | 1,518 | 1,767 | 1 | 0,124 | 1,865 | 11 | 2,193 | 500 | 1,631 | 5,265E-07 |
| Regional elections 2015 | 2,119 | 4,747E-03 | 2,160 | 1,418 | 2,923E-03 | 1,555 | 1,942 | 1 | 0,041 | 1,622 | 7 | 2,225 | 500 | 1,768 | 6,663E-07 |
| Madrid-Barcelona match | 2,408 | 5,166E-03 | 2,436 | 1,751 | 8,344E-06 | 1,753 | 1,723 | 1 | -990,155 | 37,169 | 2 | 1,706 | 500 | 1,729 | 7,164E-09 |
| HSR conflict | 1,988 | 4,346E-04 | 1,997 | 1,338 | 6,188E-04 | 1,441 | 1,746 | 1 | 1,290 | 1,424 | 36 | 2,575 | 500 | 1,611 | 6,389E-08 |
| PSOE primary | 2,438 | 3,240E-03 | 2,456 | 1,563 | 1,001E-03 | 1,624 | 1,782 | 1 | -1,084 | 2,234 | 32 | 2,251 | 500 | 1,605 | 3,774E-07 |
| FARC referendum | 2,045 | 8,271E-03 | 2,113 | 1,485 | 9,829E-04 | 1,562 | 1,795 | 1 | -1,268 | 2,228 | 7 | 1,981 | 500 | 1,561 | 1,496E-07 |
| Turkish referendum | 2,481 | 7,584E-03 | 2,516 | 1,559 | 5,781E-04 | 1,607 | 1,747 | 1 | -1,516 | 2,412 | 90 | 2,712 | 500 | 1,518 | 2,928E-08 |
| Argentinian retirement plans | 1,825 | 3,261E-03 | 1,881 | 1,553 | 7,963E-05 | 1,574 | 1,835 | 1 | -4,597 | 3,220 | 4 | 1,791 | 500 | 1,772 | 2,955E-09 |

Table S5: Parameters of each fit performed with the Wikipedia data.

| A | | | R | | | η | r | | | | |
|---|---|---|---|---|---|---|---|---|---|---|---|
| TPL | | PL | TPL | | PL | PL | Hybrid lognormal | | | PL | |
| $\alpha$ | $\lambda$ | $\gamma$ | $\alpha$ | $\lambda$ | $\gamma$ | $\gamma$ | $q$ | $\mu$ | $\sigma$ | $x_{min}$ | $\gamma$ |
| 1,521 | 1,222E-04 | 1,543 | 1,848 | 4,790E-04 | 1,864 | 2,784 | 4 | -0,606 | 0,938 | 4 | 3,966 |

17

Table S6: Values of $p_{inf}$ estimated through MLE for the Twitter datasets. The averages and standard deviations obtained for 10 samples of 1000 instances each are presented.

| Dataset | $\overline{p}_{inf}/10^{-3}$ | $\Delta p_{inf}/10^{-3}$ |
|---|---|---|
| Spanish elections 2016 | 1,56 | 0,14 |
| Spanish elections 2015 | 1,41 | 0,08 |
| Dock workers conflict | 1,05 | 0,06 |
| Palestine | 0,47 | 0,04 |
| PSOE crisis | 1,07 | 0,10 |
| Catalan elections 2015 | 3,2 | 0,4 |
| Basque elections 2016 | 1,69 | 0,14 |
| Regional elections 2015 | 1,64 | 0,09 |
| Madrid-Barcelona match | 0,23 | 0,04 |
| HSR conflict | 5,3 | 0,5 |
| PSOE primary | 0,50 | 0,05 |
| FARC referendum | 0,41 | 0,06 |
| Turkish referendum | 0,88 | 0,08 |
| Argentinian retirement plans | 0,69 | 0,07 |

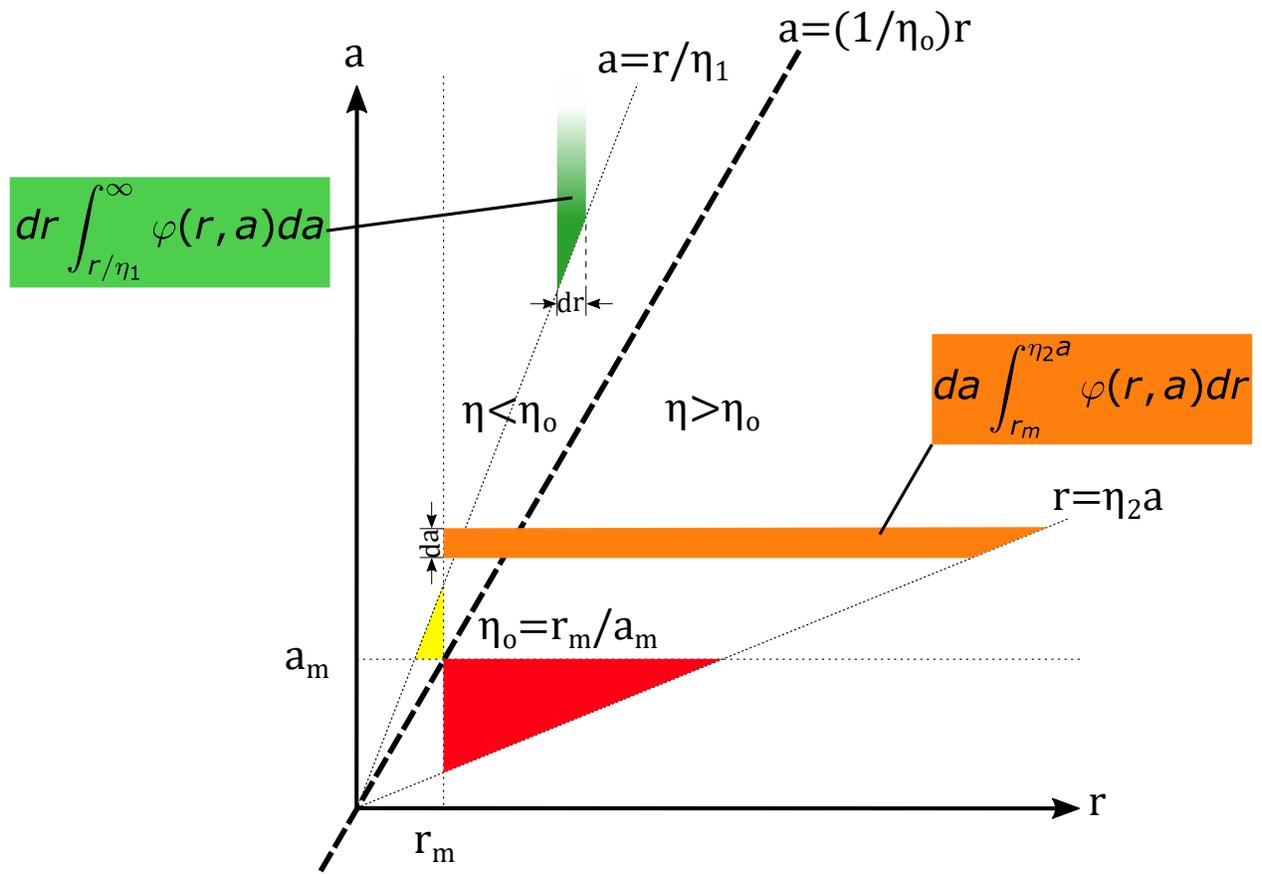

Figure S1: Diagram showing the regions of integration defined to each side of the line $\eta_o = \frac{r_m}{a_m}$ to obtain the cumulative distribution function of the efficiency.

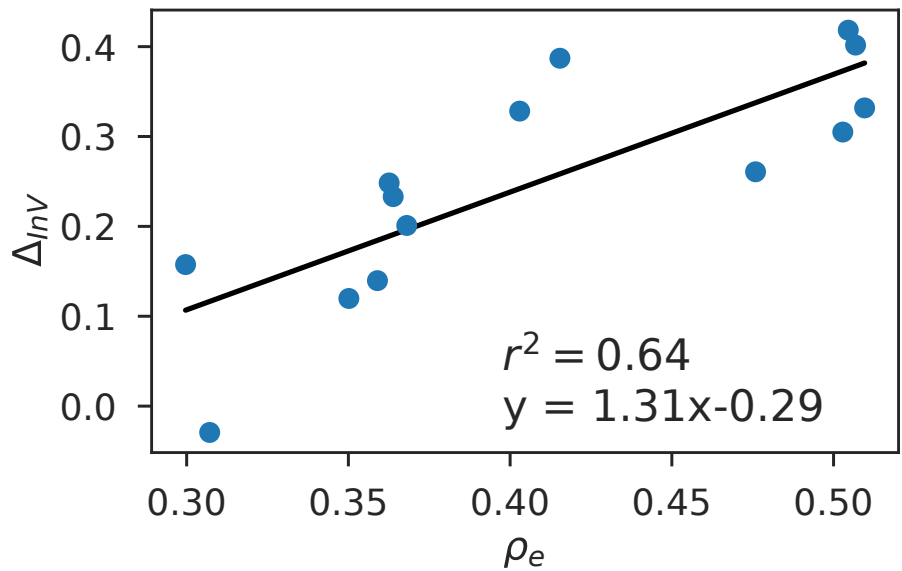

Figure S2: The deviation of the InV model ($\Delta_{InV}$) with respect to the Twitter empirical data is related to the correlations between $A$ and $R$, measured with the Spearman's correlation ($\rho_e$). The figure shows a monotonous increment of $\Delta_{InV}$ with $\rho_e$, which has been characterized by means of a linear regression.

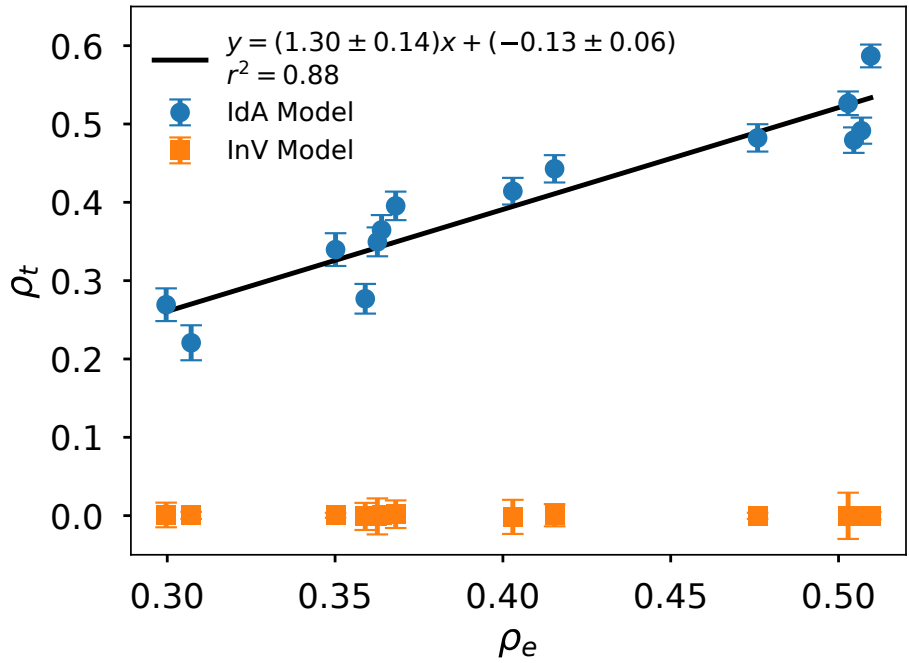

Figure S3: Linear regression of the correlation induced by the IdA model (blue dots) with respect to the empirical correlation computed for the Twitter datasets. Each point corresponds to a different conversation. For comparison the values corresponding to the InV model have also been plotted (orange squares).

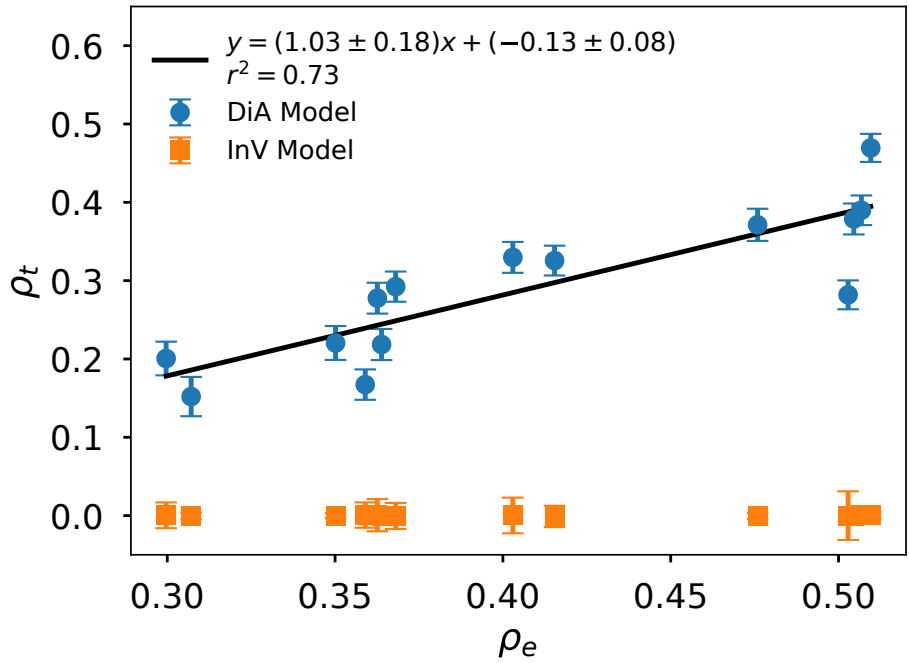

Figure S4: Linear regression of the correlation induced by the DiA model with respect to the empirical correlation computed for the Twitter datasets. Each point corresponds to a different conversation. The reference values obtained for the InV model are also displayed.

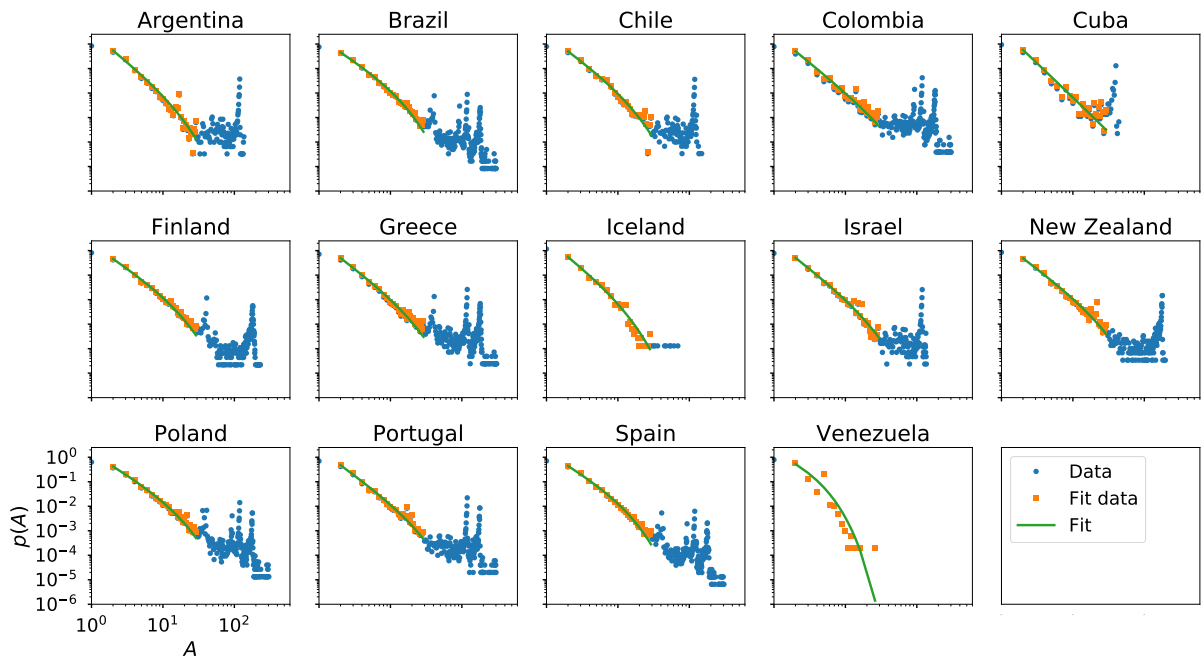

Figure S5: Fit of the empirical distribution of activity of the scientific citations network to a truncated power law. Blue dots correspond to the full data and orange squares to the data selected to perform the fit ($A \in [1, 30]$).

25

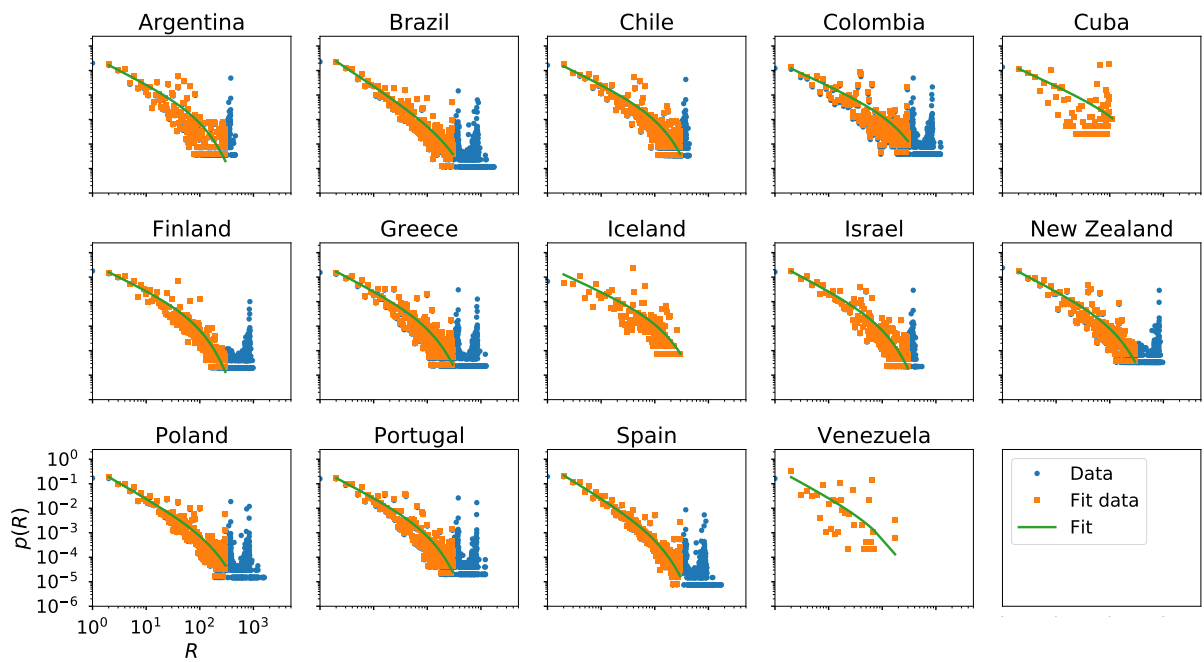

Figure S6: Fit of the empirical distribution of response of the scientific citations network to a truncated power law. Blue dots correspond to the full data and orange squares to the data selected to perform the fit ($R \in [2, 300]$).

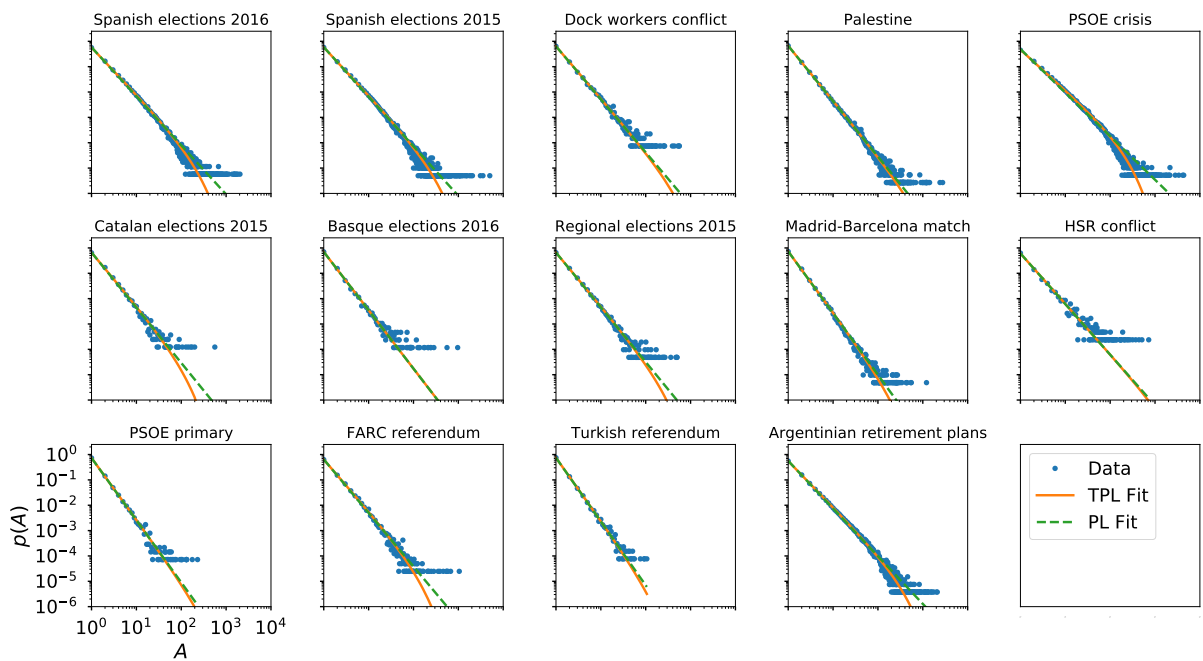

Figure S7: Fit of the empirical distribution of activity of the Twitter datasets to a power law (dashed green line) and to a truncated power law (continuous orange line).

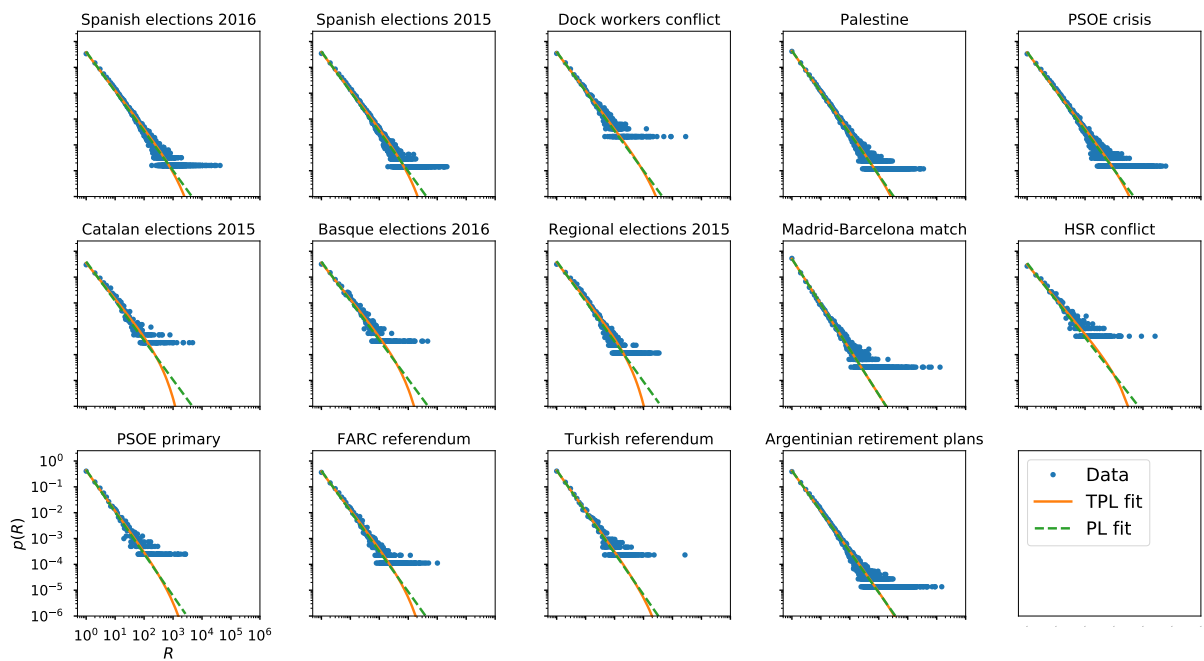

Figure S8: Fit of the empirical distribution of response of the Twitter datasets to a power law (dashed green line) and to a truncated power law (continuous orange line).

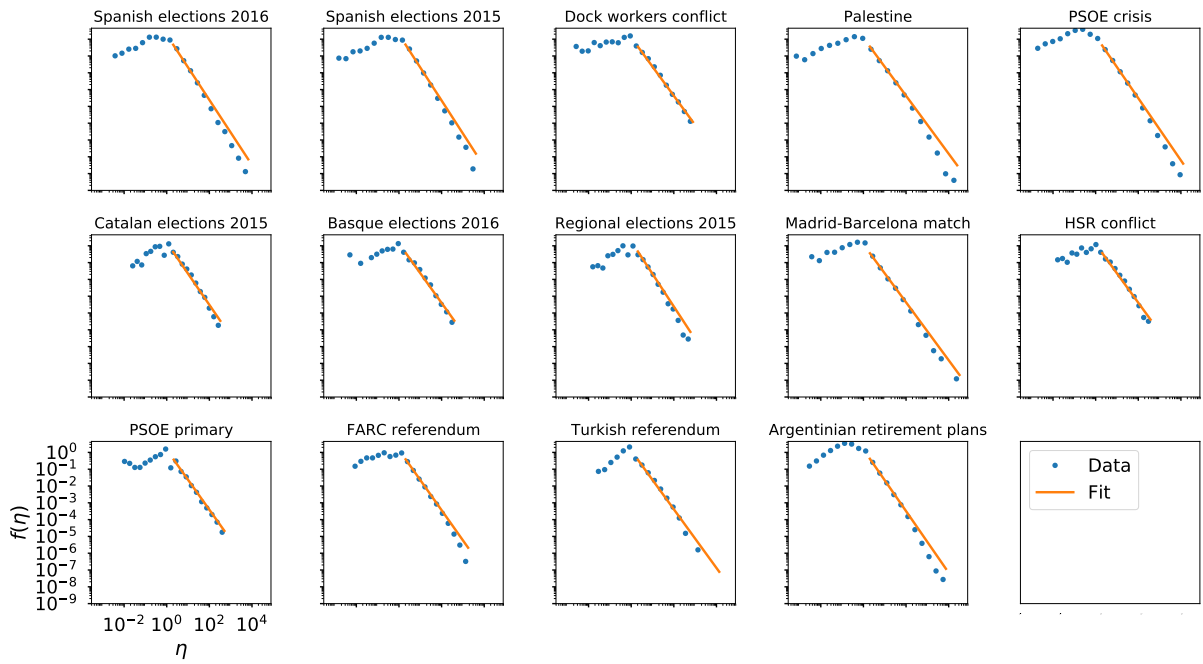

Figure S9: Fit of the right tail of the distribution of efficiency of the Twitter datasets to a power law.

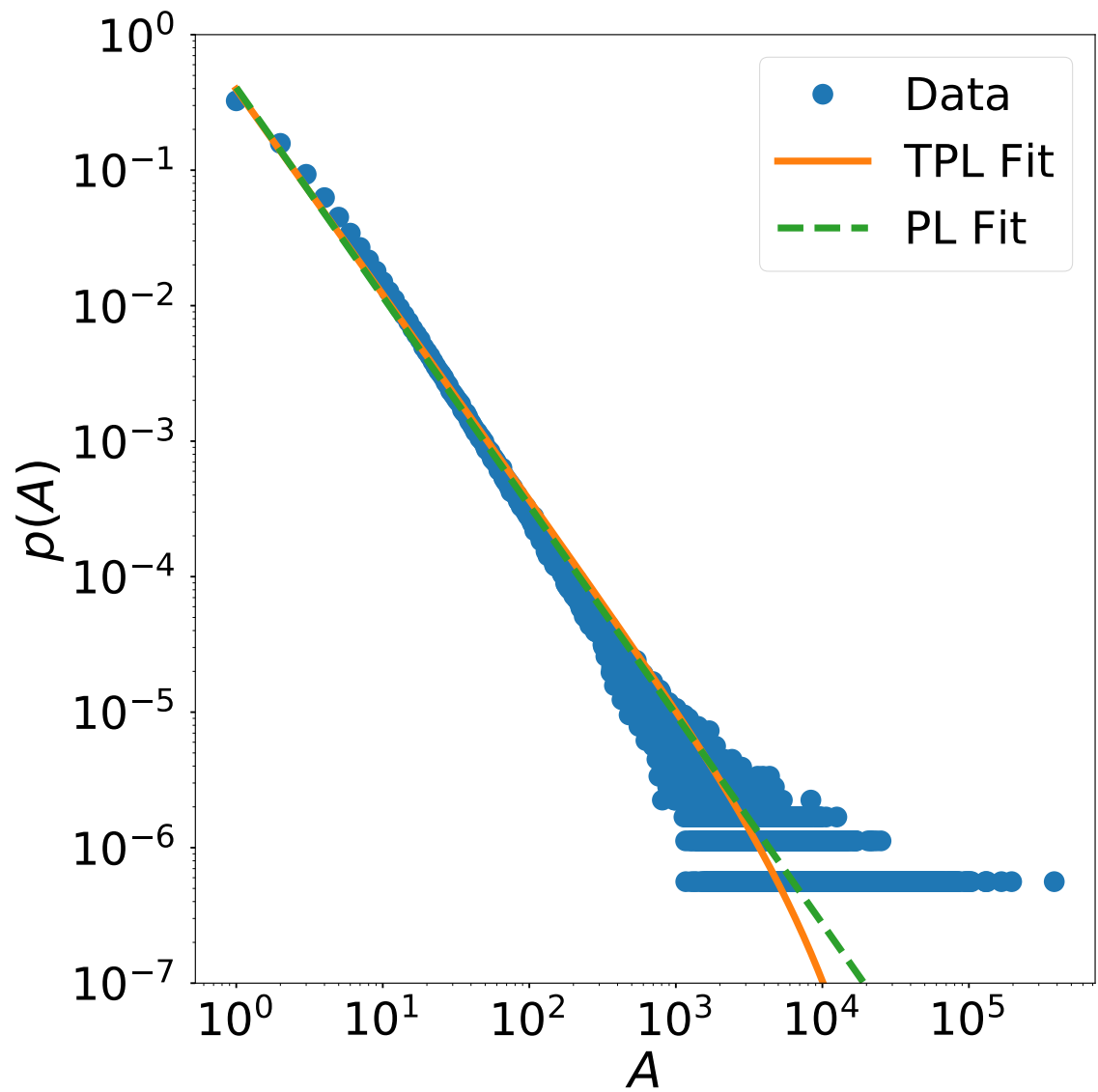

Figure S10: Fit of the empirical distribution of activity of the Wikipedia dataset to a power law (dashed green line) and to a truncated power law (continuous orange line).

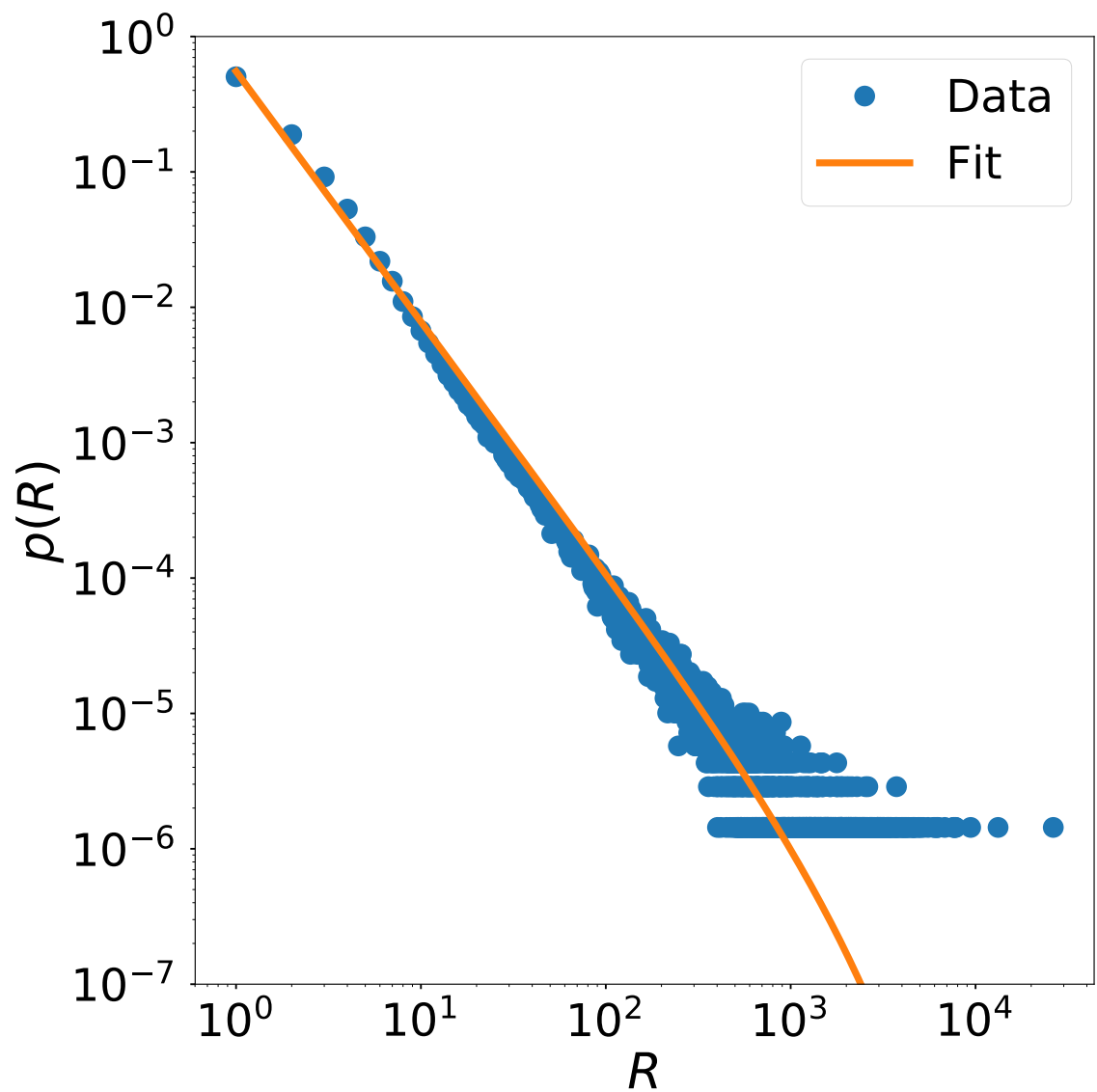

Figure S11: Fit of the empirical distribution of response of the Wikipedia dataset to a TPL.

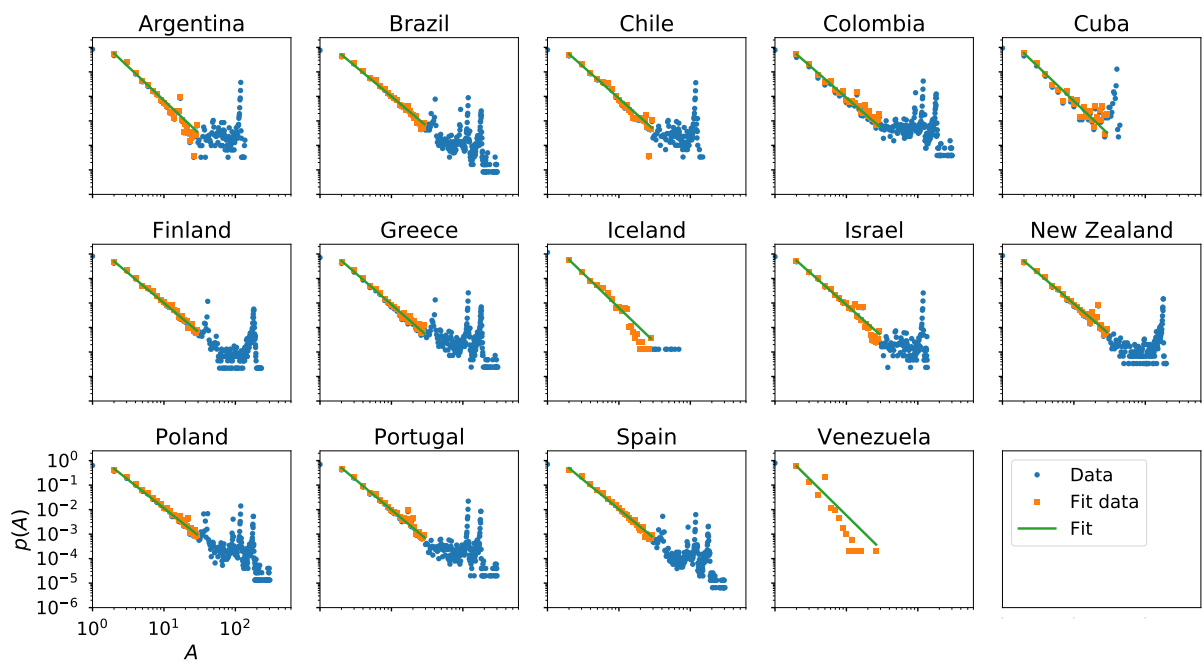

Figure S12: Fit of the empirical distribution of activity of the scientific citations network datasets to a PL.

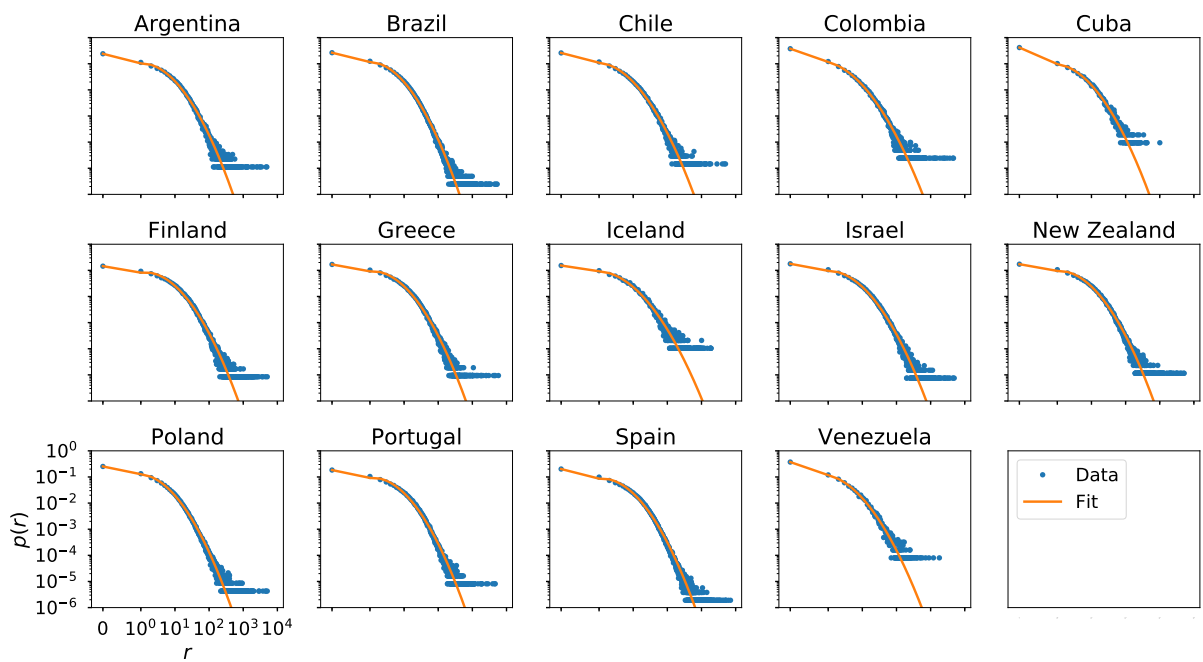

Figure S13: Hybrid lognormal fit of the empirical distribution of response to single actions of the scientific citations network datasets.

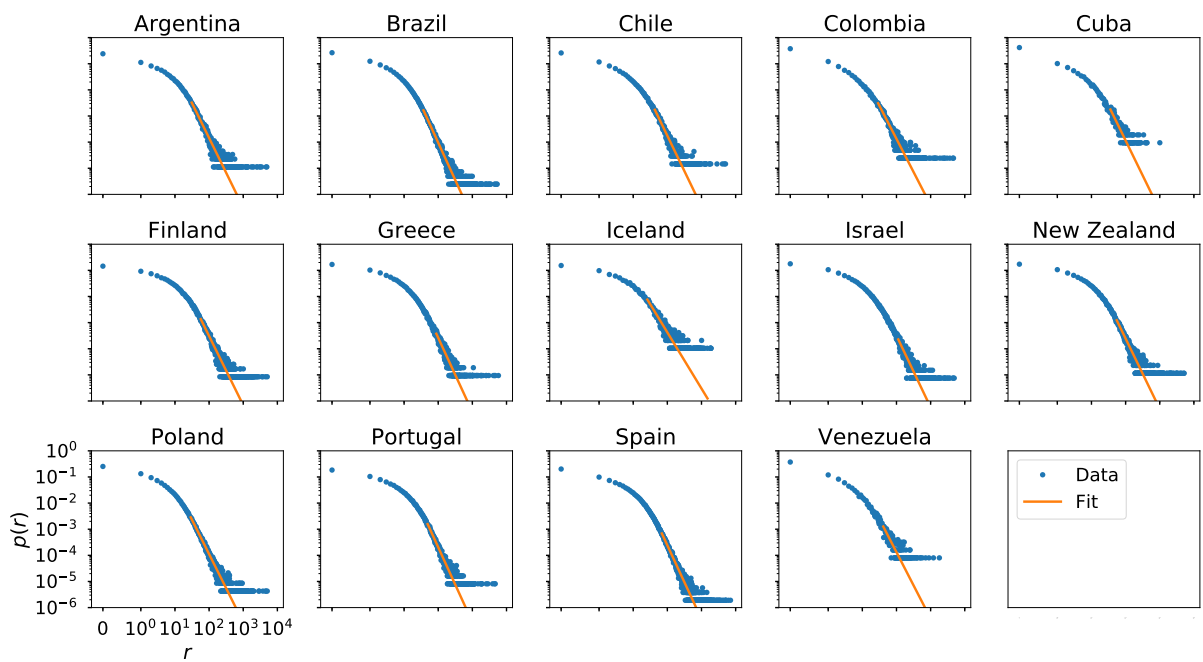

Figure S14: Power law fit of the empirical distribution of response to single actions of the scientific citations network datasets.

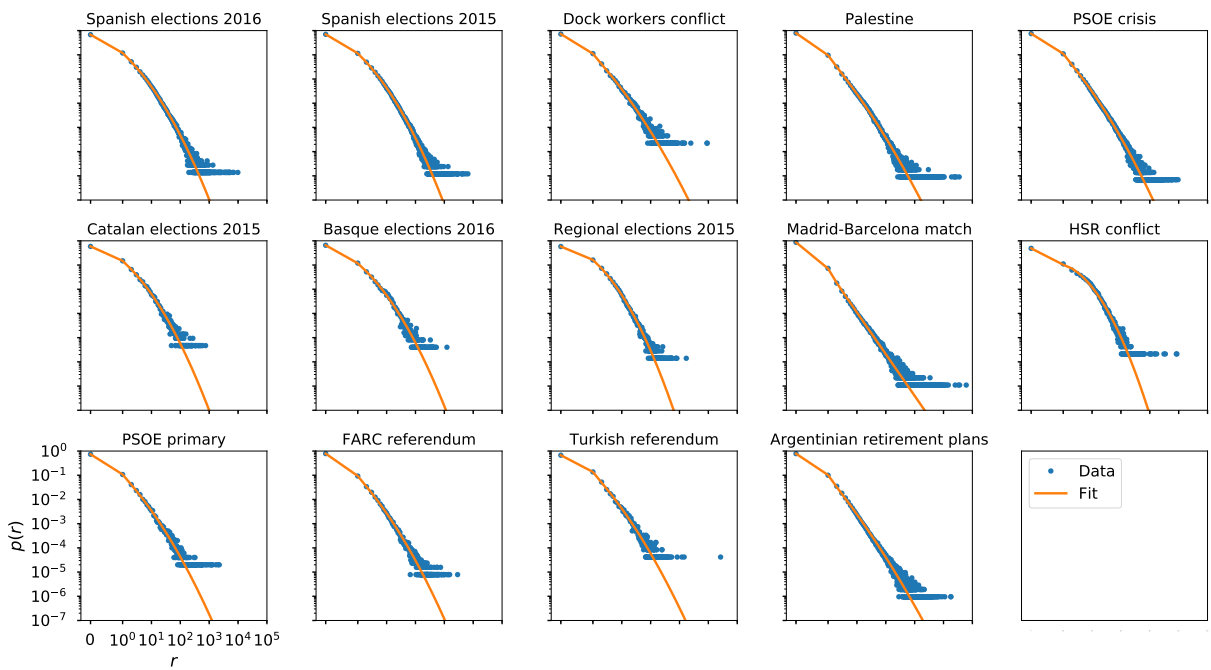

Figure S15: Hybrid lognormal fit of the empirical distribution of response to single actions of the Twitter datasets.

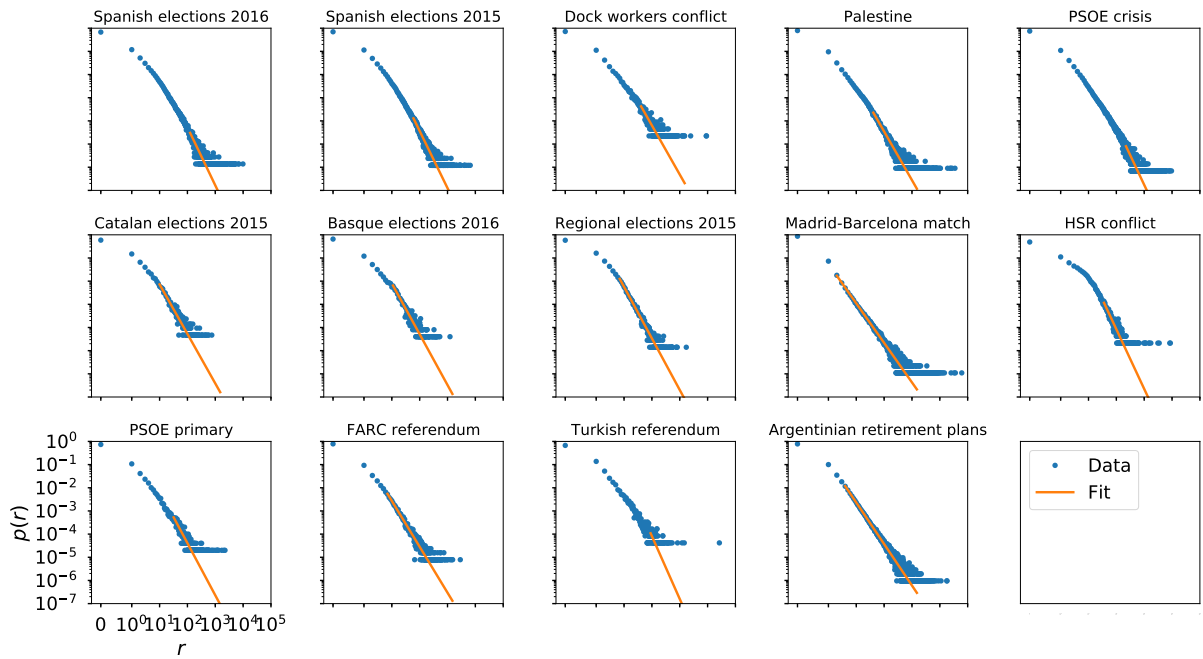

Figure S16: Power law fit of the empirical distribution of response to single actions of the Twitter datasets.

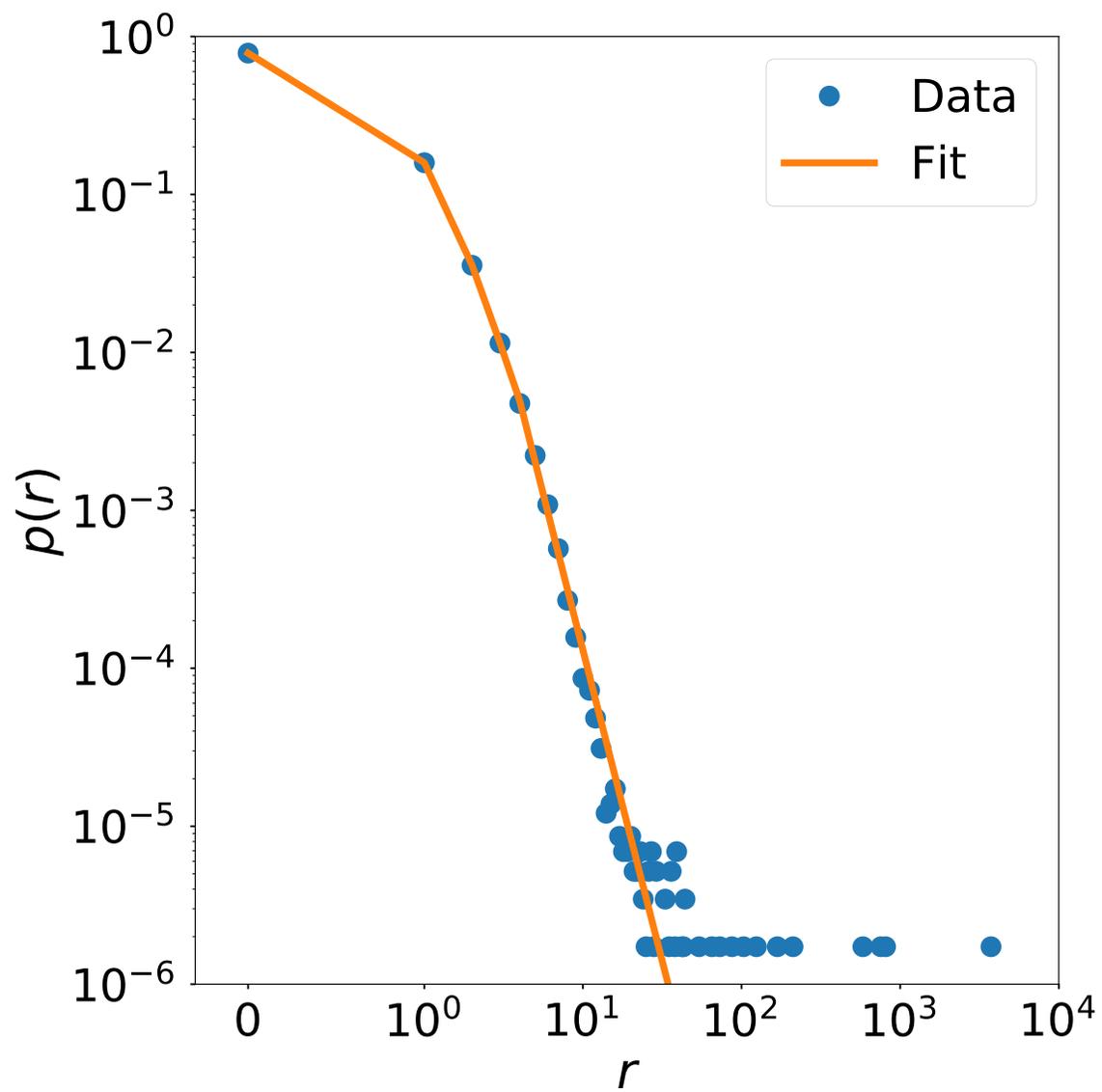

Figure S17: Hybrid PL fit of the empirical distribution of response to single actions of the Wikipedia dataset.

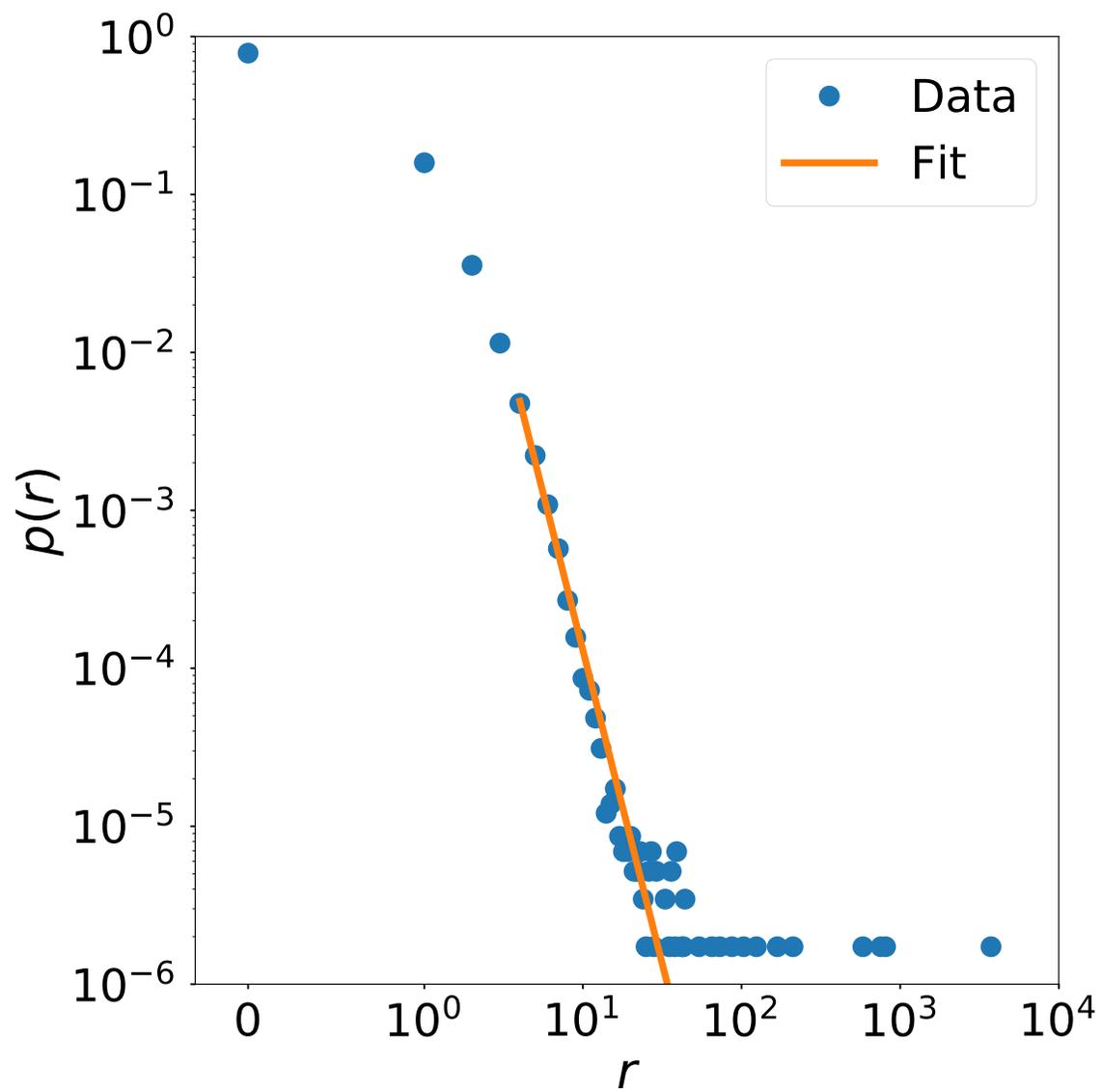

Figure S18: Power law fit of the empirical distribution of response to single actions of the Wikipedia dataset.

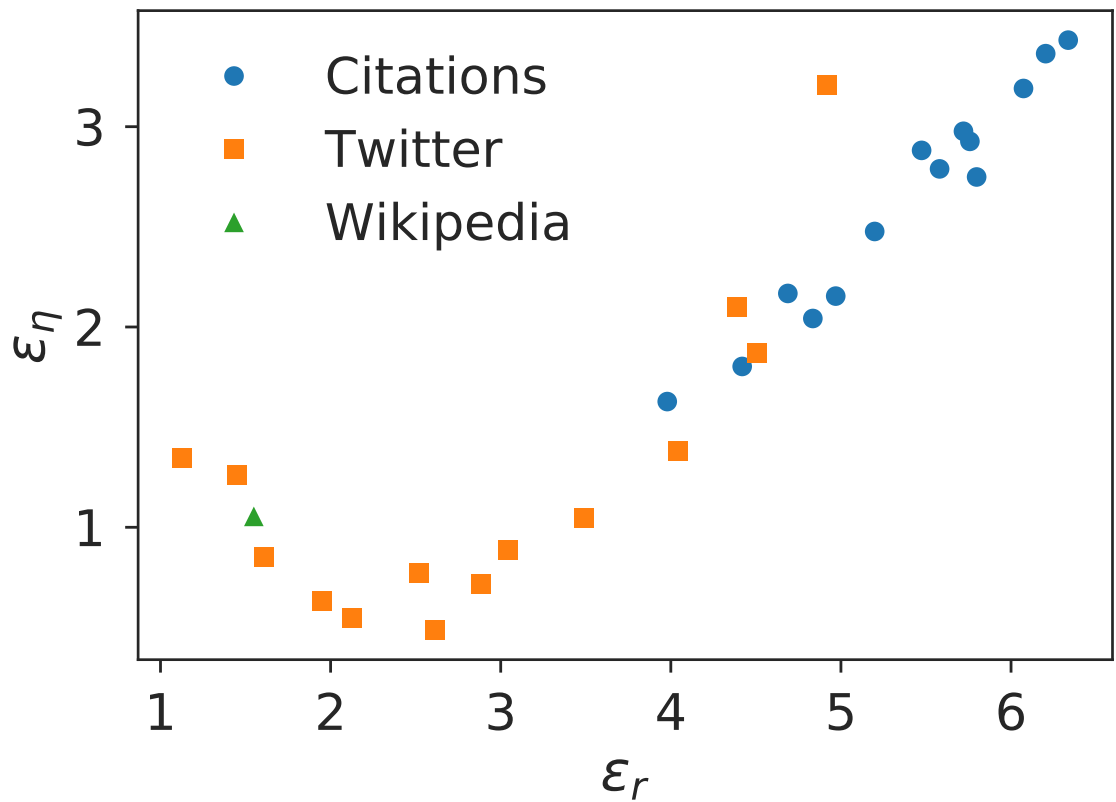

Figure S19: Relationship between the fitting error committed when $p(r)$ is modeled as a power law ($\epsilon_r$) and the deviation of the analytical approximation with respect to the numerical computation of the IdA model ($\epsilon_\eta$) in the right tail of the efficiency distribution $f(\eta)$.

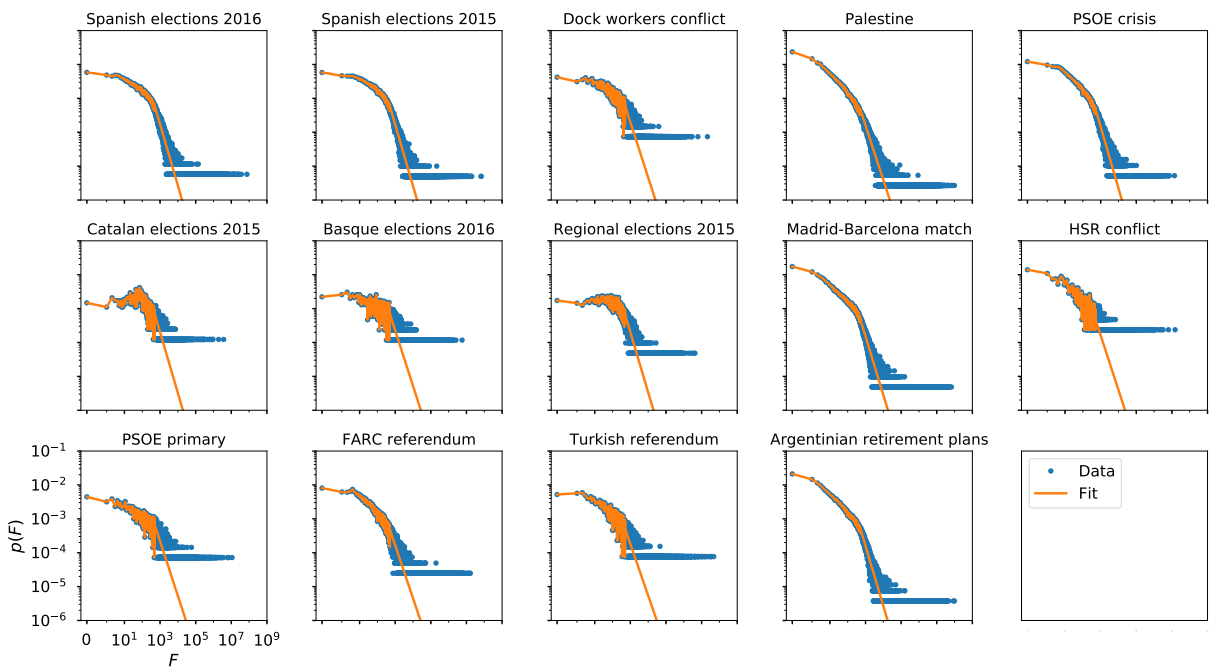

Figure S20: Fit of the empirical distribution of followers of the Twitter datasets to the empirical values with a PL tail.



\end{document}